\documentclass[lettersize,journal]{IEEEtran}
\usepackage{amsmath,amsfonts}
\usepackage{algorithmic}
\usepackage[ruled, lined, linesnumbered]{algorithm2e}
\let\oldnl\nl
\newcommand{\nonl}{\renewcommand{\nl}{\let\nl\oldnl}} 

\usepackage{array}
\usepackage[caption=false,font=normalsize,labelfont=sf,textfont=sf]{subfig}
\usepackage{textcomp}
\usepackage{stfloats}
\usepackage{url}
\usepackage{verbatim}
\usepackage{graphicx}
\usepackage{cite}
\usepackage{xcolor}

\usepackage[shortlabels]{enumitem}
\usepackage{multicol}
\usepackage{multirow}
\usepackage{tikz} 
\usetikzlibrary{positioning, shapes.geometric, backgrounds}

\newcommand\T{\rule{0pt}{2.6ex}}
\newcommand\B{\rule[-1.2ex]{0pt}{0pt}}

\usepackage[export]{adjustbox} 

\DeclareMathOperator*{\argmin}{arg\,min}

\newcounter{tmpcount}
\newcounter{continuesub}
\newlength{\tmplengthA}
\newlength{\tmplengthB}
\newlength{\tmplengthC}
\newlength{\tmplengthD}

\begin{document}

\title{Automatic Dimensionality Reduction of Twin-in-the-Loop Observers}

\author{Giacomo Delcaro, Riccardo Poli, Federico Dettù, Simone Formentin,~\IEEEmembership{{Senior Member},~IEEE},
\\
Sergio Matteo Savaresi, ~\IEEEmembership{{Senior Member},~IEEE}
\thanks{The authors are with the Electronics, Computer Science and Bioengineering Department (Dipartimento di Elettronica, Informazione e Bioingegneria), Politecnico di Milano, Milan, Italy (email: simone.formentin@polimi.it)}}

\maketitle

\begin{abstract}
Conventional vehicle dynamics estimation methods suffer from the drawback of employing independent, separately calibrated filtering modules for each variable. To address this limitation, a recent proposal introduces a unified Twin-in-the-Loop (TiL) Observer architecture. This architecture replaces the simplified control-oriented vehicle model with a full-fledged vehicle simulator (digital twin), and employs a real-time correction mechanism using a linear time-invariant output error law. Bayesian Optimization is utilized to tune the observer due to the simulator's black-box nature, leading to a high-dimensional optimization problem. This paper focuses on developing a procedure to reduce the observer's complexity by exploring both supervised and unsupervised learning approaches. The effectiveness of these strategies is validated for longitudinal and lateral vehicle dynamics using real-world data.
\end{abstract}



\section{Introduction}

\IEEEPARstart{S}{ince} the 1980s, vehicle manufacturers and tier one vendors have been developing controllers for vehicle dynamics that require unmeasured signals. Measuring these signals is often ruled out due to physical and economic constraints, therefore state observers are used to infer the missing variables. A recent survey \cite{vehicle_filters} of 33 state-of-the-art filters shows that most rely on low-order models (e.g. single-track, quarter-car) and thus estimate only a limited set of states.

\noindent
Thanks to recent advances in Hardware-in-the-Loop (HiL) platforms, it is now possible to simulate in real time a multibody Digital Twin (DT) of a vehicle onboard the vehicle itself\footnote{\label{fn:autohawk}See \url{https://www.vi-grade.com/en/products/autohawk}.}. This has enabled the integration of DTs into state estimators and control architectures, known as \textit{Twin-in-the-Loop} (TiL) systems, as demonstrated in \cite{TiL_Riva,TiL_control_long,TiL_control_lat, TiL_filtering_params, TiL_filtering_robustness,TiL_filtering_PCA}, where the authors prove the effectiveness of the approach against state-of-the-art benchmarks.

\smallskip
\noindent
This paper focuses on the TiL observer scheme, which uses the DT as model of the system and corrects the DT states in real time to track the real car. Key advantages include leveraging existing off-the-shelf vehicle simulators for modeling and the ability to estimate all vehicle states and variables simultaneously with one single observer. However, the intrinsic complexity of vehicle simulators requires treating the DT as a black-box model.

\noindent
Prior research \cite{whitebox_unified_observer, whitebox_switching_observer}, proposed a unified observer for the entire vehicle, based on full-state white-box models. However, they tend to provide lower fidelity than simulation environments when modeling complex phenomena, such as aerodynamics or suspension kinematics. This happens because simulation frameworks can leverage look‑up tables and branching logic, while control‑oriented models tend to stick to closed-form expressions, as in \cite{whitebox_unified_observer}. The TiL approach, although facing the challenge of a black-box DT, offers advantages in dealing with these complexities.

\noindent
The TiL observer's main disadvantage lies in the lack of an explicit analytical formulation of the DT, preventing the use of classical filter-tuning theories like Kalman filtering or sliding-mode-observer tuning. Instead, a black-box optimization problem is solved using Bayesian Optimization \cite{BO}, as discussed in Section \ref{subsec:parallel_bayesian_optimization}. Moreover, stability of black-box systems cannot be easily assessed: prior work \cite{TiL_filtering_robustness} has proposed probabilistic stability guarantees via scenario optimization \cite{scenario_approach} for the TiL framework, while in \cite{TiL_filtering_params} the authors randomize the initial state of the filter to check convergence dynamics.

\noindent
In the TiL Observer scheme, a correction matrix $K$ relates the output errors $e_y$ to the state corrections $e_x$ (see Figure \ref{fig:TiL_scheme}). In principle, $K$ has $n_x \times n_y$ entries---280 in our case study---making the tuning process prohibitively expensive. To keep the optimization manageable, Riva et al. \cite{TiL_Riva} restrict themselves to just five out of their original 252 parameters. Selecting which elements to tune is therefore a design choice of paramount importance in all TiL observers and remains a critical aspect in the literature.

\medskip
\noindent
\textbf{Related works.} 
This paper addresses dimensionality reduction for large-scale black-box optimization problems. Bayesian Optimization, known for effectively handling a limited number of parameters, faces practical constraints with 10 to 20 optimization variables, and in some cases as few as 5 to 6 variables, as stated in \cite{BO_frazier, BO_cod1, BO_cod2, GP_and_BO, SMGO}. 
Computational limitations in our case study lead to insufficient samples, resulting in suboptimal convergence and potential overfitting.

\noindent
To mitigate optimization complexity, one common approach is dimensionality reduction through feature selection or extraction \cite{DR_1}. Feature extraction, preferable for preserving original features, involves linear methods such as Principal Component Analysis and Factor Analysis, as well as non-linear methods like kernel linear mapping \cite{LDR, DR_1}.

\noindent
In recent years, several studies have tackled high-dimensional Bayesian Optimization, such as \cite{highdim_BO_1, highdim_BO_2, highdim_BO_3}. The underlying assumption of these papers is that the problem has low effective dimensionality, \textit{i. e.} only a subset of the problem dimensions actually influence the latent cost function. The authors adopt expansion matrices or search trees to optimize only a handful of variables in place of the original problem. In our case study, on the other hand, all variables affect the cost function: some exhibit a pronounced effect, while others have a much more modest impact. However, even less influential variables can profoundly impact the latent function by pushing the filter closer to instability. Therefore, the proposed sparse high-dimensional BO methods cannot be adopted in our case. Our objective is to identify the variables with the most significant impact on the cost function and optimize them while maintaining all other variables at zero to avoid filter instability.

\noindent
A different strategy for complexity reduction is sparse optimization (SO) \cite{SO_wright}, widely applied in fields such as compressive sensing, image processing, and matrix completion. In 2006, Candès showed that, under the Restricted Isometric Property, the NP-hard $\ell_0$-norm minimization to promote sparsity can be reformulated as a convex $\ell_1$ problem \cite{candes06}.

\noindent
While matrix completion \cite{matrix_completion} seeks to induce sparsity in matrices by minimizing rank or nuclear norm \cite{matrix_completion_with_noise}, our focus differs. Our objective is not to minimize matrix rank, but to alleviate optimization complexity by reducing matrix cardinality.

\medskip
\noindent
\textbf{Main contributions.} In the following, we will discuss two data-driven procedures to reduce the dimensionality of the optimization problem without introducing any prior to the system. This approach allows us to address the challenges of high-dimensional optimization while effectively managing the risk of instability.
\begin{enumerate}
  \item The first Dimensionality Reduction (DR) algorithm will be based on $\ell_1$-norm regularization and will be labeled \textit{Supervised DR} (SDR). This data-driven technique consists of a two-stage optimization: the first black-box optimization sorts the optimization variables based on their impact on the estimator performance.
  The second stage of this algorithm is to optimize the sole performance of the filter, using the reduced set of variables.
  \item The second DR algorithm has already been analyzed in one of our previous publications \cite{TiL_filtering_PCA}. It is based on Principal Component Analysis (PCA) and is defined \textit{Unsupervised DR} (UDR). In this work, a combined approach of SDR and UDR will be discussed.
\end{enumerate}
All the results of the previously mentioned algorithms will be compared against a state-of-the-art vehicle-state Extended Kalman Filter,  a production-grade filter and a model-based reduced version of the TiL observer.

\smallskip
\noindent
The remainder of this paper is organized as follows. The problem statement is reported in Section \ref{sec:problem_statement}. Section \ref{sec:MBR} is devoted to building a benchmark filter and a benchmark for the dimensionality reduction algorithm. Section \ref{sec:SDR} focuses on the Supervised Dimensionality Reduction (SDR) approach, and Section \ref{sec:UDR} covers the Unsupervised Dimensionality Reduction (UDR) and the combined SDR+UDR method. We conclude the paper with some final remarks in Section \ref{sec:conclusions}.


\section{PROBLEM STATEMENT}
\label{sec:problem_statement}

\subsection{TiL Architecture}
\label{subsec:TiL_architecture}
\begin{figure}[]
  \centering
  \includegraphics[width=0.9\columnwidth]{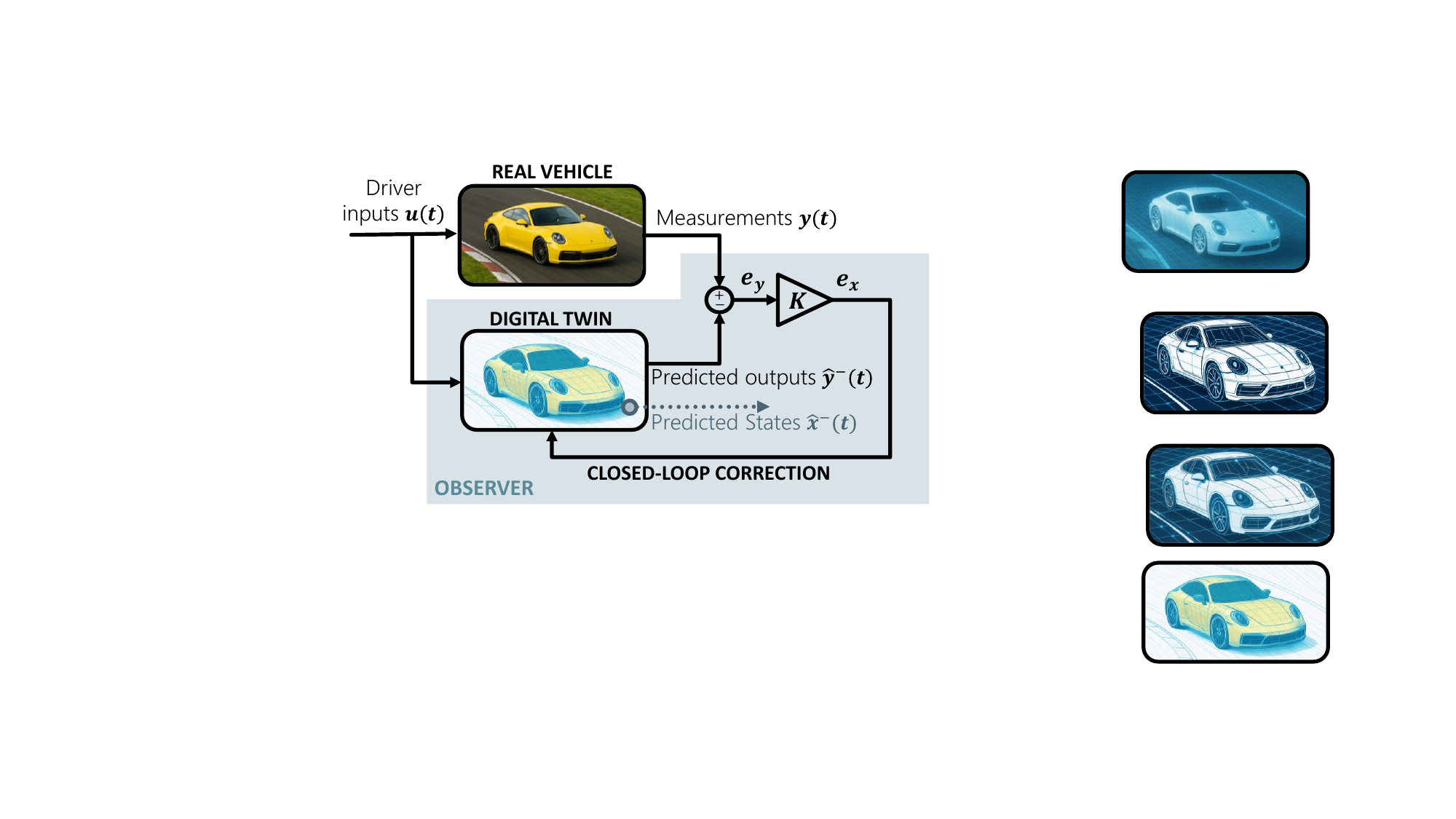}
  \caption{Twin-in-the-Loop Observer architecture.}
  \label{fig:TiL_scheme}
\end{figure}

This section describes in detail the TiL Observer architecture, reported in Figure \ref{fig:TiL_scheme}.
The scheme resembles the classical observer schemes, such as Luenberger observers or Kalman filters. The difference $e_y = y(t) - \hat{y}(t)$ between the model outputs and the measured ones is turned into a state perturbation $e_x$ via a correction matrix $K$. In discrete time this becomes a two-step procedure:
\begin{enumerate}
  \item prediction step, performed internally by the DT: 
  \begin{equation}
    \begin{cases}
      \hat{x}^-(t)\,=\,f(\hat{x}(t-1),\;u(t))\\
      \hat{y}^-(t)\,=\,h(\hat{x}(t-1),\;u(t))
    \end{cases}
    \label{eq:filter_prediction_step}
  \end{equation}
  being $\hat{x}(t)$ the estimated state, $u(t)$ the driver input and $y(t)$ the measured signal. $f(\cdot)$ and $h(\cdot)$ model the DT's behavior and are unknown to the user.
  \item correction step:
  \begin{equation}
    \hat{x}(t) = \hat{x}^-(t) + K (y(t) - \hat{y}^-(t))  \;\,.
    \label{eq:filter_correction_step}
  \end{equation}
\end{enumerate}
In Figure \ref{fig:TiL_scheme}, $e_x(t) = \hat{x}(t) - \hat{x}^-(t) $ and $e_y(t) = y(t) - \hat{y}^-(t)$. The sampling frequency of our sensors is 100 $Hz$, hence we will be correcting the DT every 10 $ms$.
\medskip\\
Since the model of the system is highly nonlinear, the extended Kalman filter framework dictates a time-varying correction law. In \cite{TiL_Riva}, the effectiveness of using just a \textit{linear, time-invariant} correction law has already been proven, hence we can adopt the same approach and optimize just one constant matrix $K$. The rationale behind this choice is the fact that the DT's outputs should be very close to the real ones, hence $e_y(t)$ will be very small, and the corrective term $e_x(t)$ will be small too (but necessary to track the real vehicle with the simulator). A time-varying matrix $K$ would not significantly improve the results, but would make the tuning problem much more complicated.

\subsection{Performance optimization}
\label{subsec:performance_optimization}
\noindent
Let us now define the procedure to tune matrix $K$ in a data-driven fashion. In order to prove the capabilities of the TiL scheme, we choose a high-performing vehicle in a racing circuit, where multiple dynamics are excited simultaneously. Moreover, in the optimization dataset the driver was asked to drive as aggressively as possible.\\
Although this paper is dedicated to laying the groundwork for data-driven TiL filter tuning rather than exhaustive scenario testing, we intentionally employ a very challenging dataset. Strong results in such a complex context generally transfer well in less demanding conditions, as shown in some of the validation datasets.

\smallskip
\noindent
In the preliminary data acquisition campaign, we need to measure both vehicle outputs $y_{meas}(t)$ and some states $x_{meas}(t)$. This is necessary since we will need to minimize the difference between the estimated states $\hat{x}(t)$ and the measured ones $x_{meas}(t)$. Note that state measurement is only necessary in the initial phase to tune $K$ and will not be required any further.

\medskip
\noindent
In order to minimize the difference between measured and estimated states, we define the following root-mean-square (rms) loss $L_K$:
\begin{equation}
  L_K(x_{meas},\hat{x}) = \sum_{i = 1}^{n_x} w_{x_i} rms(x_{{meas}_i}-\hat{x}_i) \;\,,
  \label{eq:generic_cost_function}
\end{equation}
where $w_{x_i}$ are the weights given to each state, used to normalize the different magnitudes and sort the states by importance.\\
Thus, we can minimize the estimation error by selecting $K$ such that $L_K$ is minimized. We can formulate this as an optimization problem:
\begin{subequations}
    \label{eq:generic_optimization_problem}
    \begin{gather}
        K = \argmin_{K} L_K(x_{meas},\,\hat{x})
        \label{eq:generic_optimization_problem_objective}\\
        s.t.\quad k_{i,j}\,\in\,[\,\underline{k}_{i,j},\,\overline{k}_{i,j}\,] \qquad \begin{matrix}
          \forall i = 1,\cdots,n_x\\
          \forall j = 1,\cdots,n_y
        \end{matrix}
        \label{eq:generic_optimization_problem_ranges} \;\, .
    \end{gather}
\end{subequations}
Where $k_{i,j}$ is element $(i, j)$ of matrix $K$. Equation (\ref{eq:generic_optimization_problem_ranges}) is a constraint on the range of all entries of $K$. $\overline{k}_{i,j}$ and $\underline{k}_{i,j}$ are, respectively, the upper and lower bounds for the optimization variables $k_{i,j}$.

\subsection{Parallel Bayesian Optimization}
\label{subsec:parallel_bayesian_optimization}

\noindent
Let us now select the optimization algorithm to solve problem (\ref{eq:generic_optimization_problem}). The cost function $L_K$ is, in general, non-convex and can be evaluated only by simulating the DT for each candidate $K$. Since each simulation is computationally expensive, we can explore only a limited set of matrices: the number of maximum iterations was set to $N=1000$. These requirements for minimal function evaluations and non-convex optimization lead us to adopt a sample-efficient zeroth-order global optimization algorithm \cite{zerothorderopt}: model-based (or surrogate-based) optimization (MBO) \cite{surrogate_functions_optimization}, also known as Efficient Global Optimization (EGO) \cite{ego}. At each iteration $i$, this class of algorithms fits a computationally inexpensive model of the cost function $\hat{L}_i$ and adopts this model to minimize an 
\textit{acquisition function} $a(K)$, which finds a trade-off between exploring the unknown regions and exploiting the current knowledge of the position of the minimum.\\
We will employ one of the most commonly used in this class of optimizers: Bayesian Optimization (BO) \cite{BO}. BO models $L_K$ as a Gaussian Process (GP) \cite{GP_book} with zero mean and kernel $\kappa(K,\,K')$: $L_K \sim  \mathcal{GP}(0,\;\kappa(K,\,K'))$.\\
At each iteration $i$, the GP prior is conditioned using the previously sampled data $\left\{ (K_j, \, L_{K_j}) \right\}_{j=1,...,i-1}$ to compute the GP posterior $\hat{L}_i$, which can be used as a meta-model of the unknown cost function $L_K$. This step is known as Gaussian Process Regression (GPR).\\
We will adopt one of the most frequently used kernel functions for GP regression: the Matérn $5/2$ function \cite{GP_book}:
\setcounter{tmpcount}{\value{equation}}
\begin{subequations}
  \begin{equation}
    \kappa(K_m,\,K_n) = \sigma_f^2 \left( 1+\sqrt{5}r+5r^2/3 \right)e^{-\sqrt{5}r} \;\, ,
  \label{eq:GP_matern_kernel}
  \end{equation}
  \setcounter{continuesub}{\value{equation}}\noindent
\end{subequations}
where $r$ is a measure of distance between two matrices $K_m$ and $K_n$. In high-dimensional optimization, we adopt the ARD version of the Matérn function \cite{ARD}, where every dimension is governed by its own characteristic length-scale $\ell_i$ to modulate input-space dimension smoothness:
\setcounter{equation}{\value{tmpcount}}
\begin{subequations}
  \setcounter{equation}{\value{continuesub}}
  \begin{equation}
    r = \sqrt{(K_m^{vec}-K_n^{vec})^T M (K_m^{vec}-K_n^{vec})}
  \label{eq:GP_ARD_distance}
  \end{equation}
\end{subequations}
where $M=diag([\ell_1, \cdots, \ell_d])^{-2}$ is a diagonal matrix and $K_*^{vec}$ represents the flattening of matrix $K_*$ into an $\mathbb{R}^d$ vector. The higher $\ell_i$, the smaller the impact of the $i$-th dimension on the GP. Thus, we are able to remove irrelevant dimensions.

\medskip
\noindent
As acquisition function, the expected improvement function \cite{BO_frazier} is chosen.\\
The observer gain $K_i$ sampled at iteration $i$ may not be asymptotically stable. To address this, we introduce a coupled constraint \cite{coupled_constraint}: a constraint that can be evaluated only by sampling the objective function. The constraint is modeled as an additional GP and considered in the acquisition function $a_i$, to avoid sampling points in known unstable regions (equation (8) in \cite{BO_with_constraints}).
\\
For the correct initialization, the optimization algorithm needs $n_{seed}$ initial points to start computing the meta-model. The algorithm terminates when reaching a fixed number of iterations, defined as $N$.

\medskip
\noindent
As discussed earlier, surrogate-function-based optimizers follow a sequential approach, requiring sampling and surrogate model construction to determine the next point for evaluation. To expedite this process, \textit{parallel Bayesian Optimization} \cite{parallelBO} utilizes multiple workers to sample points simultaneously. When a worker becomes available, parallel BO assigns it a new point for evaluation using all previously sampled data, including points being sampled by other workers. To expand the dataset, the algorithm employs the \textit{model-prediction} method (or kriging believer heuristic). This method imputes the value of the objective function $L(K)$ as the mean of the Gaussian Process (GP) at point $K$. To prevent overly optimistic predictions in high-dimensional models, the imputed value is saturated not to be smaller than the minimum sampled point so far, a technique known as \textit{clipped-model-prediction}.

\noindent
Parallel BO is summarized in Algorithm \ref{alg:parallel_BO}.

\begin{algorithm}[h]
    \caption{Pseudo-code for parallel BO\rule[-0.6ex]{0pt}{2.6ex}}
    \label{alg:parallel_BO}
    \SetNlSty{text}{}{:}
    \SetAlgoNlRelativeSize{-1}
    \DontPrintSemicolon
    Select $N$ number of total iterations, $n_{seed}$ number of initial points ($n_{seed}<N$) and $P$ parallel workers\;
    Sample $n_{seed}$ seed points using all workers\;
    $i \leftarrow n_{seed}+1$\;
    \While{$i \leq N$}
    {
      \If{at least one worker is free}
      {
        If any worker is busy evaluating a point, impute the value of $L_K$ in this point\;
        Using the available (actual and imputed) data, build a GP model of the objective function and the constraints\label{alg:BO_GP_step}\;
        Minimize the acquisition function $a_i(K)$: \;
        \nonl $\;\;K_i \leftarrow \argmin_K a_i(K)$\label{alg:BO_min_a_step} \;
        Evaluate $L(K_i)$ on a free worker\;
        $i \leftarrow i+1$\;
      }
    }
    Update the surrogate function\;
    \KwRet{\upshape best evaluated point}
\end{algorithm}

\subsection{Case study}
\label{subsec:case_study}

\noindent
In this case study, the proposed observer is evaluated using data acquired from a high‑performance vehicle. An initial “optimization” dataset (reported in gray in Figure \ref{fig:MBR_optimization_timeseries}) features deliberately aggressive driving---with sideslip angles higher than 23 degrees at speeds above 100 $km/h$---and serves to tune the observer by maximally exciting all dynamic modes. Once the observer is calibrated on the aggressive optimization run, its performance is then evaluated on multiple validation datasets. The primary validation dataset (Validation A), reported in gray in Figure \ref{fig:MBR_validation_timeseries}, involves a full‑lap run with high but more consistent driving demands. Four additional validation scenarios (datasets B–E), covering a range of driving conditions, are introduced in Section \ref{sec:MBR} to test the generality of the results.

\medskip
\noindent
The available sensors are:
\begin{itemize}
  \item 6-dof IMU (3D accelerometer and gyroscope) (6 signals);
  \item wheel encoders (4 signals).
\end{itemize}
A Real-Time-Kinematic positioning system (RTK-GPS) was installed on the vehicle to supply ground-truth measurements for our experiments. Note that the GPS data will not be used as input to the observer, since it was mounted for testing purposes only. Besides, in typical driving scenarios GPS sensors suffer from availability and signal quality problems, caused for example by tall buildings or tunnels.

\medskip
\noindent
Due to the absence of suspension data, the vehicle's vertical dynamics will not be analyzed and the DT will only track the real longitudinal and lateral motions. To achieve this, the cost function in (\ref{eq:generic_cost_function}) must weigh states which influence both dynamics. It is straightforward to verify that without one of the two, the DT will not be able to track the real vehicle. The cost function is then defined as follows:
\begin{equation}
  L_K(x_{meas},\,\hat{x})=\underbrace{rms(v_x-\hat{v}_x)}_{\text{long. dynamic}}+\underbrace{rms(\omega_z-\hat{\omega}_z)}_{\text{lat. dynamic}}
  \label{eq:actual_cost_fcn}
\end{equation}
Where the yaw rate $\omega_z$ is expressed in $deg/s$ while the longitudinal velocity $v_x$ is in $km/h$. The first term in $v_x$ is necessary to track the longitudinal dynamic of the real vehicle, while $\omega_z$ the lateral one. Only one state per dynamic is used to keep the cost function as simple as possible.

\medskip
\noindent
Since the DT must support real-time execution on an ECU and allow the external correction of its internal states, we employ VI-Grade's VI-CarRealTime (CRT)\footnote{https://www.vi-grade.com/en/products/vi-carrealtime/}. CRT is a commercially available, industry‑proven black‑box vehicle simulator capable of modeling complex phenomena, such as engine 3D suspension dynamics, combustion-engine and powertrain power delivery, and discontinuous tire-road contact behaviors. As shown in Figure 18 of \cite{TiL_control_long}, CRT can execute onboard an AutoHawk ECU (see footnote \ref{fn:autohawk}), operating at 1 $kHz$ and consuming approximately 40\% of the computational power of one of its cores.\\
To simulate the vehicle, CRT adopts 28 internal states:
\begin{itemize}
  \item 3D position, orientation, linear and angular velocities;
  \item wheels' positions and velocities;
  \item suspensions' stroke lengths and velocities.
\end{itemize}
Since CRT can run on Simulink, we will use MATLAB function \texttt{bayesopt}.

\medskip
\noindent
Correcting all 28 states with all 10 measured signals would generate a correction matrix with $n_x \times n_y = 280$ entries, which is far too highly dimensional for BO and cannot be tuned effectively in $N=1000$ iterations. Imposing sparsity in the problem while maintaining the number of iterations can be beneficial: high-dimensional search spaces may prevent the optimizer from reaching the global optimum, while a reduced number of variables is more likely to converge. However, since the large-scale formulation includes all variables, its optimum cannot be inferior to that of the reduced problem, hence the latter may have better convergence but to a worse solution.

\medskip
\noindent
In this paper, we limit our analysis to a subset of the system’s states and outputs. In particular, since no GPS data is available, we will not correct the vehicle's pose or its orientation. Since the front and rear wheels are highly coupled, only the velocity of one wheel for each pair will be corrected. No data from the suspensions is available, hence the vehicle's vertical dynamics will not be considered. For the same reason, roll and pitch rates and vertical velocity will not be corrected. Finally, for the lateral dynamic we will only weigh $\omega_z$, since it is directly measured and is involved in the cost function equation (\ref{eq:actual_cost_fcn}).\\
To sum up, we will only correct the following 4 states:
\vspace*{-0.75\multicolsep}
\begin{multicols}{2}
\begin{itemize}
    \item $v_x$: longitudinal velocity
    \item $\omega_z$: yaw rate
    \item $\omega_{fl}$: FL wheel speed
    \item $\omega_{rr}$: RR wheel speed
\end{itemize}
\end{multicols}
\noindent
Finally, we need to select the outputs used to correct the aforementioned states. Since three of the four states are directly measured, we will use only the following outputs:
\vspace*{-1.6\multicolsep}
\begin{multicols}{2}
\begin{itemize}
    \item $\omega_z$: yaw rate
    \item $\omega_{fl}$: FL wheel speed
    \item $\omega_{rr}$: RR wheel speed
\end{itemize}
\end{multicols}

\noindent
The cardinality of the problem is therefore $n_x \times n_y = 12$. Optimizing such a number of parameters with a surrogate-function-based optimizer probably will not converge to the true optimum, it nonetheless brings the problem within computational reach. We will now discuss how to tackle such a problem, and how reducing its dimensionality affects the performance, convergence, and computational times. The aim of this work does not reside in the pursuit of the best possible observer or the estimation of the most complex dynamics, which was already proven in \cite{TiL_Riva}, but rather in formulating methodologies for the reduction of dimensionality of the filter.


\section{Building the Benchmarks}

\noindent
Prior to presenting our data-driven reduction approaches (Sections \ref{sec:SDR} and \ref{sec:UDR}), we establish two benchmarks. Section \ref{sec:ekf_benchmark} develops an EKF reference model to serve as performance benchmark for every TiL observer. Section \ref{sec:MBR} introduces instead a benchmark for the dimensionality reduction, using expert domain knowledge to prune unnecessary entries in the TiL correction matrix.

\subsection{Benchmark: dynamic single-track EKF}
\label{sec:ekf_benchmark}

\noindent
Several recent surveys have offered a comprehensive overview of vehicle state estimation techniques, revealing prevailing trends and widely adopted practices across both research and industry. Among these studies, such as those presented in \cite{survey_Puscul} and \cite{survey_Chindamo}, the Extended Kalman Filter (EKF) consistently emerges as the most established approach for onboard vehicle state estimation. Its ability to handle nonlinear vehicle dynamics through successive linearizations, along with its moderate computational requirements, make it a robust and practical choice for onboard implementation; however, it requires analytic Jacobians and risks divergence if the nonlinearities become too strong or the update rate too low.\\
Different vehicle models can be employed within the EKF framework, ranging from simple kinematic formulations to more detailed dynamic models. Among these, dynamic models are generally preferred for their ability to better capture tire and inertial effects, which are critical for accurate estimation of vehicle states.
In particular, we take as reference the EKF-based estimator proposed by Reina et al. \cite{ekf_reina}, which builds upon the widely used dynamic bicycle model. The corresponding state vector includes:
\[
\mathbf{x} = 
\begin{bmatrix}
v_x & v_y & \omega_z & F_x & F_{y,f} & F_{y,r}
\end{bmatrix}^T
\]
where \(v_x\) and \(v_y\) denote the longitudinal and lateral velocities, \(\omega_z\) the yaw rate, \(F_x\) the longitudinal force and \(F_{y,f|r}\) the lateral tire forces at the front and rear. The only input to the model is the steering angle \(\delta\), while the process noise is modeled as zero-mean Gaussian white noise. The parameters of the vehicle dynamic model---such as mass, yaw inertia, axle distances, drag and road-tyre rolling resistance coefficients---were determined through identification procedures based on experimental data.\\
The adopted measurements from the onboard sensors are:
\[
\mathbf{y} = 
\begin{bmatrix}
a_x & a_y & \omega_z & v_x
\end{bmatrix}^T
\]
where \(a_x\) and \(a_y\) are the longitudinal and lateral accelerations, and \(v_x\) is indirectly computed from wheel-speed sensors.
The tuning of the EKF was carried out by optimizing the process noise covariance matrix \(Q\) and the measurement noise covariance matrix \(R\) over the optimization dataset. The optimal values were obtained by minimizing a cost function defined as the rms error between the estimated and reference values of the longitudinal velocity, the sideslip angle and the yaw rate.\\
The results for the optimization and validation (A) datasets are reported in Figures \ref{fig:MBR_optimization_timeseries} and \ref{fig:MBR_validation_boxplot} (yellow line), demonstrating accurate estimation of both $v_x$ and $\omega_z$. Further analysis of these results are reported in the next Sections.

\medskip
\noindent
As a further benchmark, we also evaluate an OEM filter provided by an industrial partner. This benchmark is a bank of single‐dynamic observers, each meticulously calibrated for a particular dynamic and driving condition. This strategy exemplifies the industry practice of splitting complex dynamics into separate, meticulously calibrated observers, at the cost of a considerable calibration effort.\\
The results for the OEM filters for the optimization and validation (A) datasets are reported in Figures \ref{fig:MBR_optimization_timeseries} and \ref{fig:MBR_validation_boxplot} (black dashed lines).

\subsection{Benchmark: model-based reduction}
\label{sec:MBR}

\noindent
Having presented the EKF and OEM baselines, we now introduce the TiL dimensionality-reduction benchmark. In our case study (Section \ref{subsec:case_study}), the correction matrix $K$ has 12 entries, where each element $k_{y_j\to x_i}$ indicates that state $x_i$ is corrected using the error on the measurement $y_j$.\\
We exploit physics‐based insights to remove redundant optimization variables, defining this approach as Model-Based (or Physics-Inspired) Reduction (MBR). By ranking each entry of $K$ according to its importance, we can impose different sparsity levels and thus study how aggressive pruning affects overall performance. The most important parameters are the auto-correlated ones (\ref{list:param_importance1}) (lower diagonal of $K$). The second and third class of parameters are respectively made of the parameters that relate the wheel speeds to the longitudinal velocity (\ref{list:param_importance2}) and cross-correlate the wheel speeds (\ref{list:param_importance3}). The last class relates lateral and longitudinal dynamics (\ref{list:param_importance4}).
\vspace*{-0.75\multicolsep}
\begin{multicols}{3}
\begin{enumerate}[label=\roman*), ref={\roman*}, noitemsep, topsep=0pt]
    \item $k_{\omega_z\to\omega_z}$ \addtocounter{enumi}{-1} \label{list:param_importance1}
    \item $k_{\omega_{fl}\to\omega_{fl}}$ \addtocounter{enumi}{-1}
    \item $k_{\omega_{rr}\to\omega_{rr}}$ 
    \item $k_{\omega_{fl}\to v_x}$ \addtocounter{enumi}{-1} \label{list:param_importance2}
    \item $k_{\omega_{rr}\to v_x}$
    \item $k_{\omega_{fl}\to \omega_{rr}}$ \addtocounter{enumi}{-1} \label{list:param_importance3}
    \item $k_{\omega_{rr}\to \omega_{fl}}$
    \item $k_{\omega_z\to v_x}$ \addtocounter{enumi}{-1} \label{list:param_importance4}
    \item $k_{\omega_z\to \omega_{fl}}$ \addtocounter{enumi}{-1}
    \item $k_{\omega_z\to \omega_{rr}}$ \addtocounter{enumi}{-1}
    \item $k_{\omega_{fl}\to \omega_z}$ \addtocounter{enumi}{-1}
    \item $k_{\omega_{rr}\to \omega_z}$ \addtocounter{enumi}{-1}
\end{enumerate}
\end{multicols}

\label{subsec:optimization_ranges_heuristic}
\noindent
\textbf{Optimization ranges heuristic.}
Once the optimization parameter have been selected, problem (\ref{eq:generic_optimization_problem}) must be solved over its prescribed bounds. Although often overlooked, range selection in zeroth-order methods can be critical: a well-designed heuristic can reduce the number of evaluations and improve the solution quality. We categorize parameters into three groups:

\begin{enumerate}
  \item Auto-correlated parameters: $k_{\omega_z\to\omega_z}$, $k_{\omega_{fl}\to\omega_{fl}}$ and $k_{\omega_{rr}\to\omega_{rr}}$. Since the dynamics of the system are quite slower than the DT update frequency (100 $Hz$), we can safely state that the states can be approximated as slowly changing:  $x_i(t+1) \approx x_i(t)$. The predictor (eq. \ref{eq:filter_prediction_step} and \ref{eq:filter_correction_step}) becomes: $\hat{x}_i(t) \approx \hat{x}_i(t-1) + k_{i \to i}(x_i(t)-\hat{x}_i(t-1))$, which is asymptotically stable for $k_{i \to i}\in(0,\;2)$. In our tests, gains near 2 caused filter instability, so we conservatively limit to $k_{i \to i}\in(0,\;1.5)$. 
  \item Cross-correlated wheel parameters: $k_{\omega_{fl}\to\omega_{rr}}$ and $k_{\omega_{rr}\to\omega_{fl}}$. They share the same nature of $k_{i \to i}$, thus we will keep the span of 1.5 but center it at zero.
  \item The intervals of the remaining parameters are roughly given by range($x_i$)/range($y_j$)\footnote{The range $x(t)$ is defined as: $\text{range}(x(t))=\max_t(x(t))-\min_t(x(t))$\T}.  Therefore, the data was cut in equally long intervals and the span of each variable was set to $1.5\cdot\text{mean}(\text{range}(x_i)/\text{range}(y_j))$. All intervals are centered at zero, except for $k_{\omega_{fl}\to v_x}$ and $k_{\omega_{rr}\to v_x}$, which are nonnegative.
\end{enumerate}
The optimization variable ranges are reported in Table \ref{table:variable_ranges}.

\begin{table}[]
  \centering
  \begin{minipage}{.45\columnwidth}
    \centering
    \begin{tabular}{| c | c | c |}
      \hline
      \textbf{Parameter} & $\boldsymbol{\underline{k}}$ & $\boldsymbol{\overline{k}}$\T\B \\
      \hline \hline
      \textbf{$k_{\omega_z\to v_x}$} & -1.5 & 1.5 \T \\
      \textbf{$k_{\omega_z\to\omega_z}$} & 0 & 1.5 \T \\
      \textbf{$k_{\omega_z\to \omega_{fl}}$} & -1.5 & 1.5 \T \\
      \textbf{$k_{\omega_z\to \omega_{rr}}$} & -1.5 & 1.5 \T \\
      \textbf{$k_{\omega_{fl}\to v_x}$} & 0 & 0.45 \T \\
      \textbf{$k_{\omega_{fl}\to \omega_z}$} & -0.06 & 0.06 \T\B \\
      \hline
    \end{tabular}
  \end{minipage}
  \hspace{-0.2cm}
  \begin{minipage}{.45\columnwidth}
    \centering
    \begin{tabular}{| c | c | c |}
      \hline
      \textbf{Parameter} & $\boldsymbol{\underline{k}}$ & $\boldsymbol{\overline{k}}$\T\B \\
      \hline \hline
      \textbf{$k_{\omega_{fl}\to\omega_{fl}}$} & 0 & 1.5 \T \\
      \textbf{$k_{\omega_{fl}\to \omega_{rr}}$} & -0.75 & 0.75 \T \\
      \textbf{$k_{\omega_{rr}\to v_x}$} & 0 & 0.45 \T \\
      \textbf{$k_{\omega_{rr}\to \omega_z}$} & -0.06 & 0.06\T \\
      \textbf{$k_{\omega_{rr}\to \omega_{fl}}$} & -0.75 & 0.75 \T \\
      \textbf{$k_{\omega_{rr}\to\omega_{rr}}$} & 0 & 1.5 \T\B \\
      \hline
    \end{tabular}
  \end{minipage} 
  \\[5pt]
  \caption{Optimization variables ranges.}
  \label{table:variable_ranges}
  \vspace{-0.5cm}
\end{table}

\begin{figure}[]
  \centering
  \includegraphics[width=0.95\columnwidth]{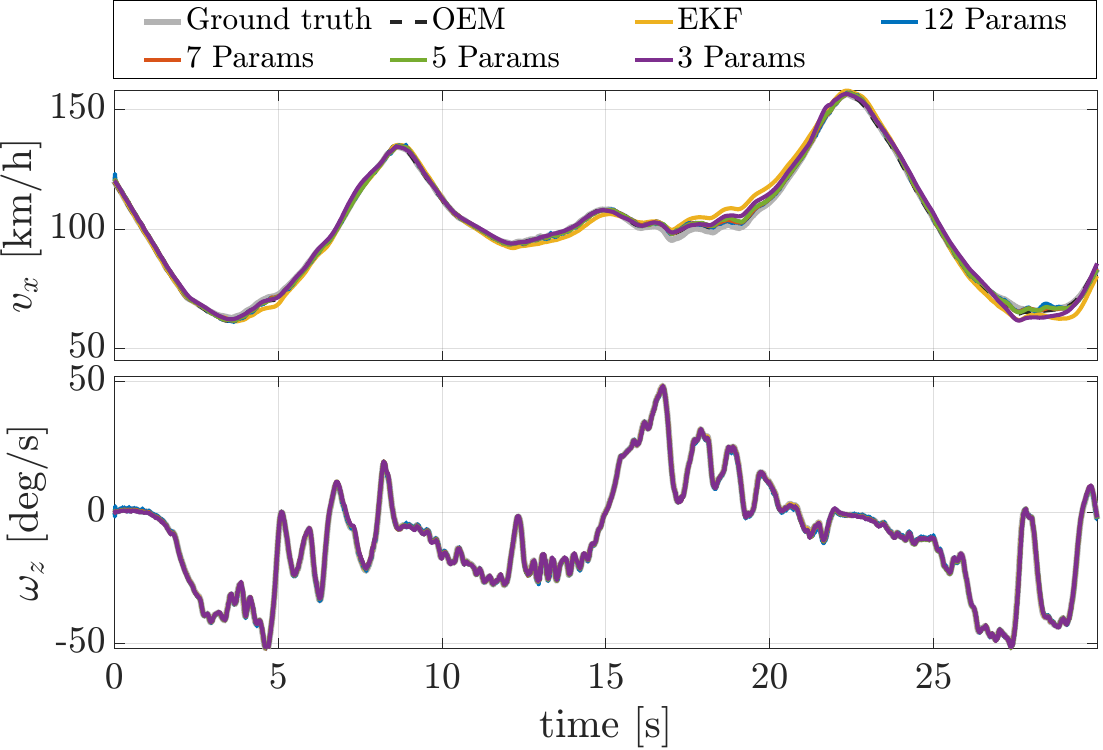}
  \vspace{-0.25cm}
  \caption{MBR: Filter performance comparison in the optimization dataset. TiL Filters with different numbers of parameters (12 to 3) are compared against the ground truth (grey line), EKF and OEM.}
  \label{fig:MBR_optimization_timeseries}
\end{figure}

\begin{figure}[]
  \centering
  \includegraphics[width=0.95\columnwidth]{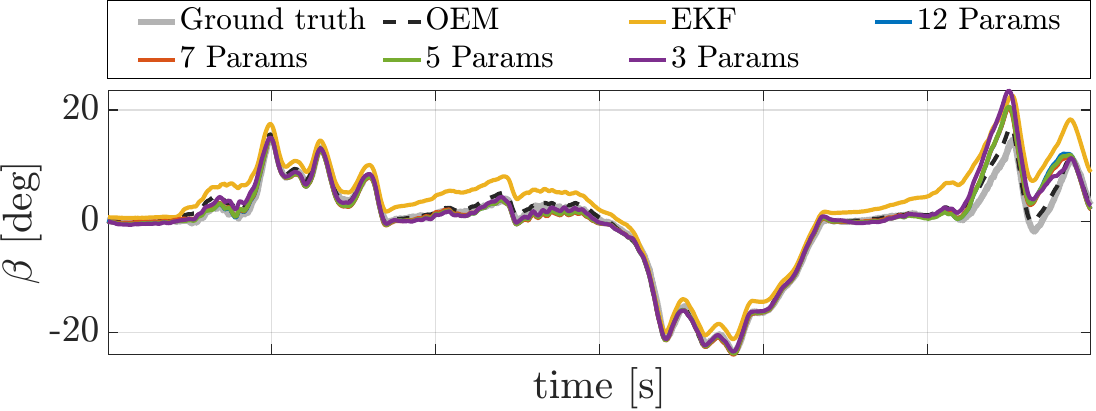}
  \vspace{-0.25cm}
  \caption{MBR: $\beta$ estimation comparison in the optimization dataset.}
  \label{fig:MBR_optimization_timeseries_beta}
\end{figure}

\medskip
\noindent
\textbf{Optimization stack.}
\label{subsec:MBR_optimization_stack}
The performance optimization problem becomes:
\begin{subequations}
  \label{eq:optimization_problem}
  \begin{align}
    K = &\argmin_{K} \; rms(v_x-\hat{v}_x) + rms(\omega_z-\hat{\omega}_z)
    \label{eq:optimization_problem_objective}\\
    s.t.
    \quad &k_{i,j}\,\in\,[\,\underline{k}_{i,j},\,\overline{k}_{i,j}\,] \qquad
    \begin{matrix}
      \forall i = 1,\cdots,n_x\\
      \forall j = 1,\cdots,n_y
    \end{matrix}
    \label{eq:optimization_problem_ranges}\\
    &\mathcal{G}(K) \leq 0
    \label{eq:optimization_problem_stability}
  \end{align}
\end{subequations}
Where constraint (\ref{eq:optimization_problem_stability}) represents the set of $K$ for which the filter is asymptotically stable. $\mathcal{G}(K)$ is unknown and must be estimated alongside the objective function: it is a coupled constraint. A simulation is deemed unstable if its state trajectories diverge.
\\
The problem is solved via Parallel BO (Algorithm \ref{alg:parallel_BO}), with $P=2$ workers, $n_{seed}=500$ and $N=1000$ feasible iterations---i.e., only evaluations that yield stable filters.
\medskip\\
Four variants of the optimization are performed, according to the parameter importance in (i)-(iv): the full 12-parameter case, and reduced problems with 7, 5, and 3 parameters. The results of the optimizations are reported in Figure \ref{fig:MBR_optimization_timeseries}. The state estimates of the different TiL filter reductions are compared with the ground truth, the OEM filters and the EKF of section \ref{sec:ekf_benchmark}. The plots represent longitudinal velocity $v_x$ (top) and yaw rate $\omega_z$ (bottom), which are weighted in the cost function (\ref{eq:optimization_problem}). A rigorous statistical assessment of each variable will be presented later.\\
Moreover, Figure \ref{fig:MBR_optimization_timeseries_beta} shows that even the unoptimized lateral sideslip state $\beta$---one of the most demanding tasks in vehicle state estimation---is tracked with high accuracy by the DT, while the EKF shows higher errors.

\begin{figure}[]
  \centering
  \includegraphics[width=0.95\columnwidth]{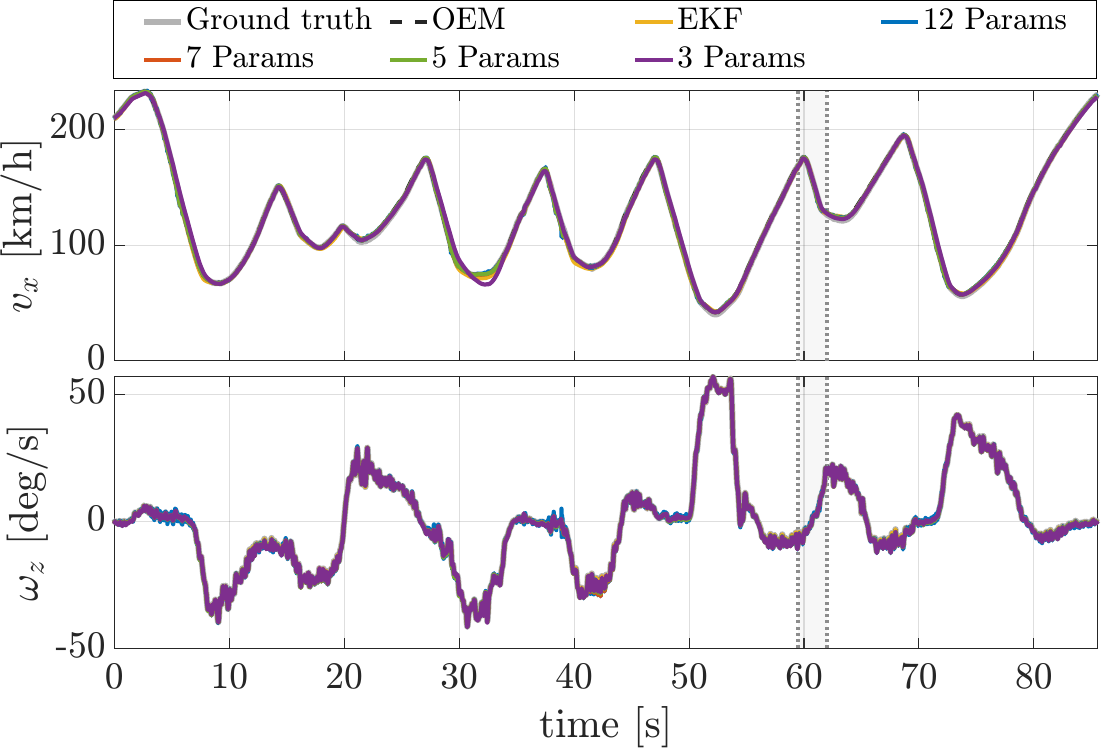}
  \vspace{-0.25cm}
  \caption{MBR: Filter performance comparison in the validation dataset.}
  \label{fig:MBR_validation_timeseries}
\end{figure}

\begin{figure}[]
  \centering
  \includegraphics[width=0.95\columnwidth]{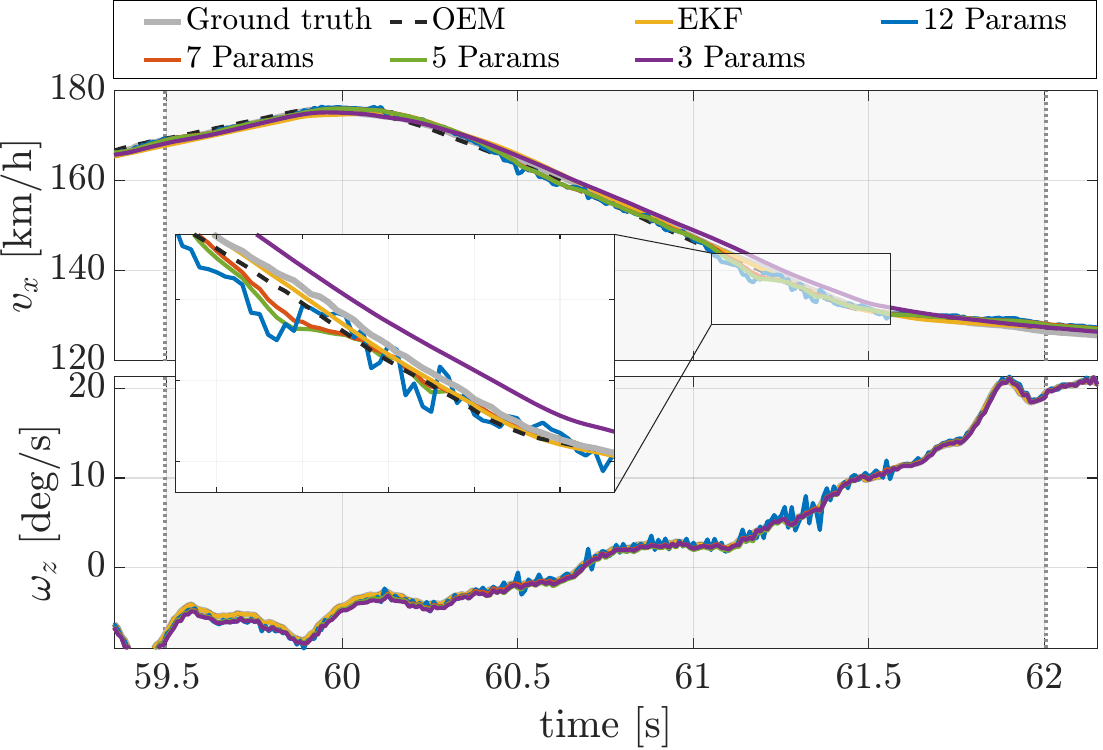}
  \vspace{-0.25cm}
  \caption{MBR: Zoomed segment of Figure \ref{fig:MBR_validation_timeseries}.}
  \label{fig:MBR_validation_timeseries_zoomed}
\end{figure}

\noindent
The validation dataset (dataset A) is reported in Figure \ref{fig:MBR_validation_timeseries}. The filters seem to be able to generalize quite well and do not overfit the optimization dataset. A zoomed segment of the validation dataset is shown in Figure \ref{fig:MBR_validation_timeseries_zoomed}. It can be noted that the 12-parameter optimization presents more noise than the lower-dimensional ones, thus it has a higher estimation error. As the parameter count decreases, the estimated signals become progressively smoother; the 3-parameter solution, however, is over-smoothed and loses critical detail.

\begin{figure}[]
  \centering
  \includegraphics[width=0.95\columnwidth]{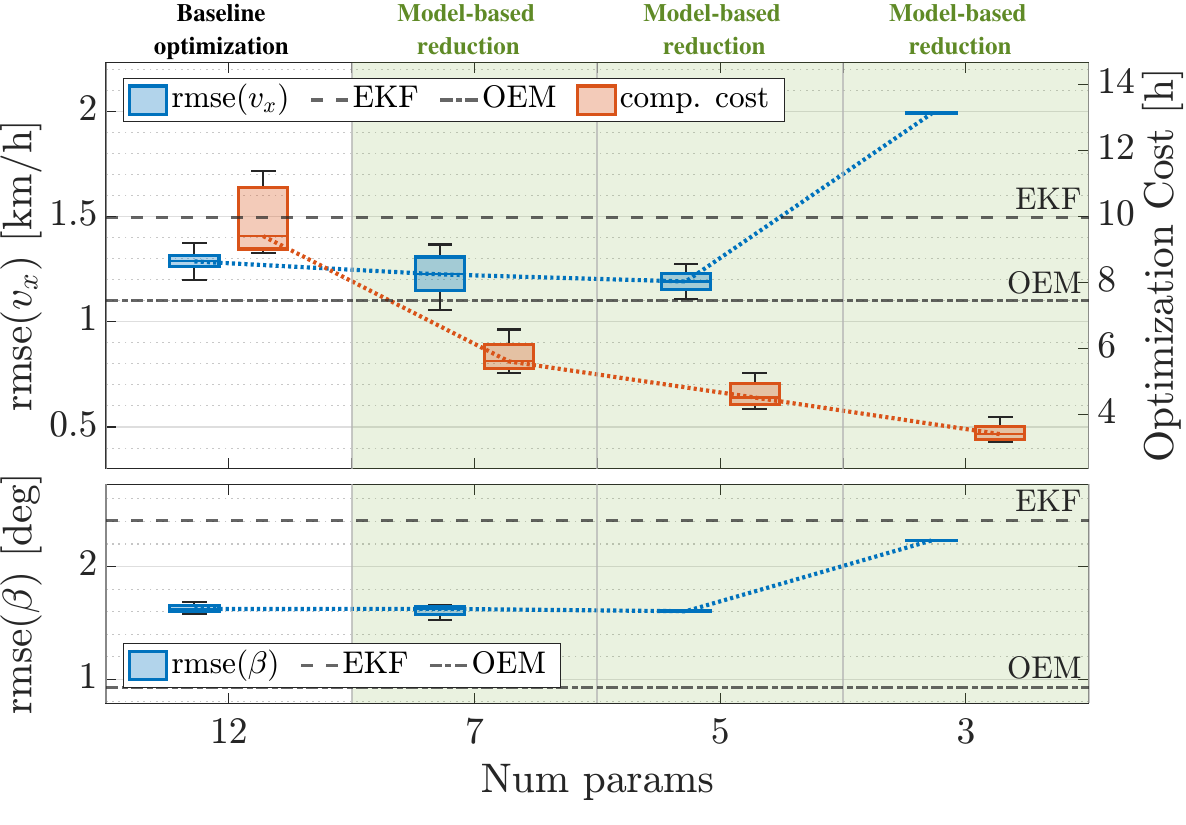}
  \vspace{-0.25cm}
  \caption{MBR: Validation A boxplot. Blue boxes report the distributions of rmse. Orange boxes represent the computational effort. Optimizations are sorted by descending dimensionality.}
  \label{fig:MBR_validation_boxplot}
\end{figure}

\medskip
\noindent \textbf{Statistical performance analysis.} To evaluate statistical consistency, each optimization was executed 12 times. All optimal solutions remained well within the prescribed bounds, except for the 3-parameter case, where its three auto-correlated gains saturated at the upper limit of 1.5.
\medskip\\
The statistical results for the validation lap for all the TiL observers are reported in Figure \ref{fig:MBR_validation_boxplot}. The top panel reports the RMSE of $v_x$ (blue boxes, left axis) and computational cost (orange boxes, right axis), while the bottom panel shows the RMSE of the sideslip angle $\beta$. EKF and OEM performance are included for comparison. The boxes represent the 25\textsuperscript{th} and 75\textsuperscript{th} percentiles of the distributions, while the horizontal black line represents the median. The whiskers illustrate the minimum and maximum values. The cost $rmse(\omega_z)$ is not reported since $\omega_z$ is directly measured and only weighted in the objective function to enforce lateral dynamic consistency; its tracking performance is not of primary interest.\\
Figure \ref{fig:MBR_validation_boxplot} shows that reducing the number of gains from 12 to 5 slightly lowers the loss, but further pruning to 3 gains increases it, highlighting that the full problem is too large to converge reliably, yet an overly sparse model lacks necessary flexibility. Moreover, the variance tends to decrease as the dimensionality is reduced. This is due to the better convergence to the global minimum for the simplified problems. When comparing TiL, OEM, and EKF, the OEM filter delivers superior accuracy on $v_x$ and $\beta$, reflecting its bank of finely-tuned single-dynamic observers. The TiL observer consistently outperforms the EKF but does not match the OEM’s precision. Remarkably, even without explicit tuning for $\beta$ in the TiL optimization problem (\ref{eq:optimization_problem}), the integrated full-vehicle DT allows more accurate sideslip estimation than the EKF.

\medskip
\noindent
\textbf{Computational effort.} As shown in Figure \ref{fig:MBR_validation_boxplot}, the computational effort increase with the problem dimensionality. At each iteration $i$, fitting the GP meta‐model with $d$ input dimensions has a complexity of $\mathcal{O}(i^2d+i^3)$ \cite{GP_book}, accumulating to $\mathcal{O}(N^3d+N^4)$ over $N$ iterations. Following model construction, the minimization of the nonconvex acquisition function also becomes more demanding as $d$ increases, since the solver must explore a higher-dimensional search space. Furthermore, higher-dimensional problems explore in general 10 to 20\% more unfeasible points, increasing even more the overall optimization effort. These combined effects explain the monotonic increase in optimization time as dimensionality rises.

\medskip
\noindent
\textbf{Additional validations.} To verify that the observed benefits of dimensionality reduction are not specific to a single lap, we introduce four supplementary validation sets (B–E), each comprising 90–110 s of driving with 6–8 turns and intervening straights. These scenarios differ substantially from the original optimization and validation laps:
\begin{itemize}
    \item B: predominantly low‑acceleration maneuvers with milder corners and lighter throttle/braking, yielding reduced longitudinal and lateral excitation;
    \item C: high throttle/brake activity on straights but slow turns, exciting only longitudinal dynamics;
    \item D: comprises prolonged, high‑intensity combined longitudinal and lateral maneuvers, exceeding the ones in the optimization dataset;
    \item E: emphasis on lateral dynamics, with pronounced cornering inputs but moderate longitudinal accelerations.
\end{itemize}
Taken together, B–E encompass scenarios spanning milder to more aggressive dynamics relative to the optimization dataset.

\begin{figure}[]
  \centering
  \hspace*{-0.08\columnwidth}
  \includegraphics[width=0.88\columnwidth]{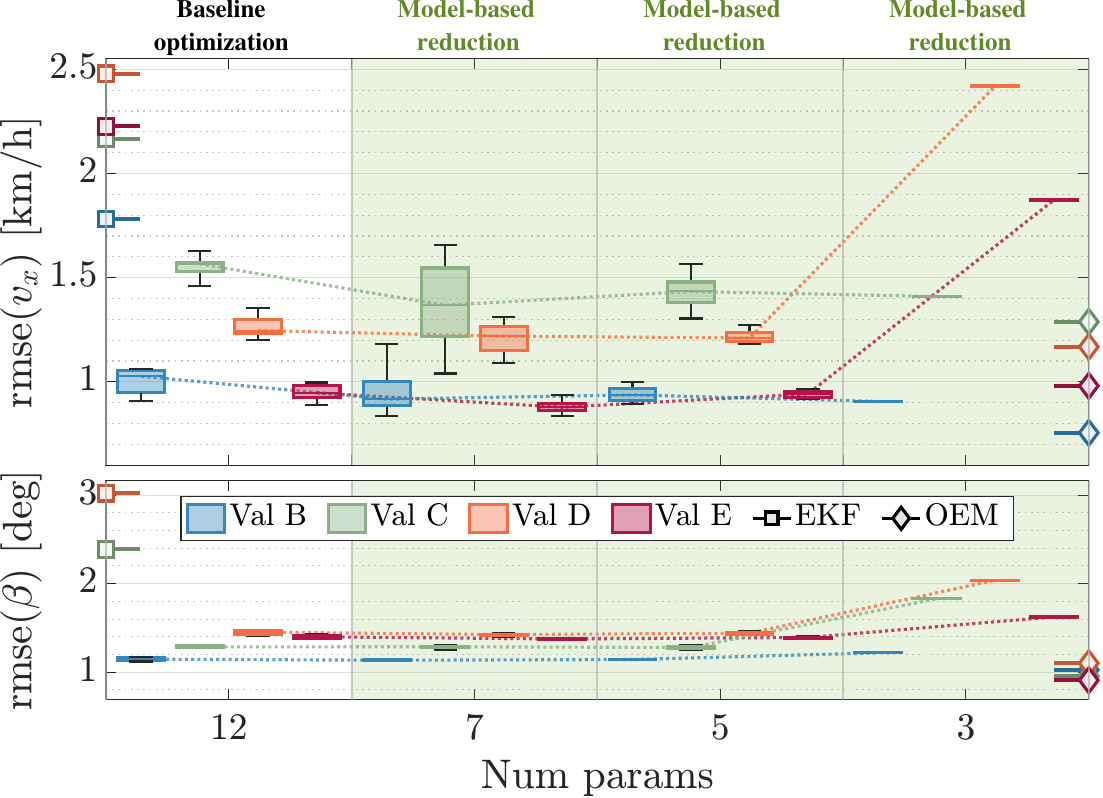}
  \vspace{-0.25cm}
  \caption{MBR: Additional validations boxplot for multiple datasets.}
  \label{fig:MBR_extraval_boxplot}
\end{figure}

\noindent
Figure \ref{fig:MBR_extraval_boxplot} shows $v_x$ and $\beta$ for the validations B-E. In all datasets, the reduced-dimensionality TiL observers produce a slight but consistent improvement in median over the 12-dimensional baseline, confirming the same trends of Figure \ref{fig:MBR_validation_boxplot}, accompanied by a similar contraction in variability. Although the absolute RMS values differ across datasets---reflecting their unique driving scenarios---the error’s dependence on dimensionality follows the same pattern in every dataset.
Remarkably, all TiL observers outperform the state‑of‑the‑art EKF---marked by dataset‑colored squares, with $\beta$ values for B and E exceeding the plot upper limit---yet they do not exhibit a statistically significant advantage over the OEM filters---marked by dataset-colored diamonds. In our tests, adjusting $Q$ and $R$ yielded lap‑specific improvements at the expense of performance elsewhere. Dataset‑targeted tuning of these matrices produced better results only for that dataset and fell short of surpassing TiL or OEM.

\medskip
\noindent
In summary, MBR TiL observers consistently outperform the EKF and approach the accuracy of OEM filters, while also estimating unoptimized states effectively thanks to the integrated full‑fledged DT. Unlike the OEM bank of single-dynamic filters, the TiL framework leverages the simulated model to maintain fidelity across all states. Reducing the observer’s dimensionality yields a substantial reduction in computational cost without degrading---and in fact slightly improving---filter performance and result convergence.

\section{Supervised dimensionality reduction}
\label{sec:SDR}

\noindent
This section introduces Supervised Dimensionality Reduction (SDR), a data-driven method to reduce problem dimensionality without prior variable-importance knowledge, unlike MBR.

\subsection{Sparsity requirement}
\label{subsec:sparsity_requirement}

\noindent
Sparse Optimization (SO) focuses on finding sparse solutions to optimization problems \cite{SO, SO_wright}, in particular for undetermined linear systems. Although the usual linear‐system assumptions do not apply in our setting, the fundamental principles of SO remain applicable. Three canonical sparse problem formulations follow:
\setlength{\tmplengthA}{\abovedisplayskip}
\setlength{\tmplengthB}{\abovedisplayshortskip}
\setlength{\tmplengthC}{\belowdisplayskip}
\setlength{\tmplengthD}{\belowdisplayshortskip}
\setlength{\abovedisplayskip}{2 pt}
\setlength{\abovedisplayshortskip}{2 pt}
\setlength{\belowdisplayskip}{2 pt}
\setlength{\belowdisplayshortskip}{2 pt}
\begin{itemize}[topsep=0pt]
  \item $\ell_0$-constrained problem formulation:
  \setcounter{tmpcount}{\value{equation}} 
  \begin{subequations}
    \label{eq:SO}
    \begin{equation} 
      K = \argmin_{K} L_K(x_{meas},\,\hat{x}) \quad s.t. \;\; \| K \|_0 \leq T \; ;
      \label{eq:SO_l0}
    \end{equation}
    \setcounter{continuesub}{\value{equation}}
  \end{subequations}
  \item function-constrained formulation:
  \setcounter{equation}{\value{tmpcount}} 
  \begin{subequations}
    \setcounter{equation}{\value{continuesub}}
    \begin{equation}      
      K = \argmin_{K} \| K \|_0 \quad s.t. \;\; L_K(x_{meas},\,\hat{x}) \leq \overline{L}  \; ;
      \label{eq:SO_fc}
    \end{equation}
    \setcounter{continuesub}{\value{equation}}
  \end{subequations}
  \item weighted formulation:
  \setcounter{equation}{\value{tmpcount}} 
  \begin{subequations}
    \setcounter{equation}{\value{continuesub}}
    \begin{equation}      
      K = \argmin_{K} L_K(x_{meas},\,\hat{x}) + \lambda \| K \|_0  \; .
      \label{eq:SO_weighted}
    \end{equation}
  \end{subequations}
\end{itemize}
\setlength{\abovedisplayskip}{\tmplengthA}
\setlength{\abovedisplayshortskip}{\tmplengthB}
\setlength{\belowdisplayskip}{\tmplengthC}
\setlength{\belowdisplayshortskip}{\tmplengthD}
Where $\| K \|_0$ is the $\ell_0$-norm of matrix $K$: $\| K \|_0 = \sum_{i,j} \mathbf{1}_{\{\,k_{ij}\neq 0\,\}}$. $T$ in equation (\ref{eq:SO_l0}) is the maximum allowed number of parameters different from zero and $\overline{L}$ in equation (\ref{eq:SO_fc}) is an upper bound on the cost. $\lambda$ in equation (\ref{eq:SO_weighted}) is a hyperparameter that---by Lagrange multiplier theorem---can be set so that the formulation equals (\ref{eq:SO_l0}) or (\ref{eq:SO_fc}).
\medskip\\
The $\ell_0$-norm in Eqs. (\ref{eq:SO}) introduces discontinuities that make the problem unsuitable for BO. We therefore adopt the $\ell_1$-norm minimization used in SO to restore tractability.
\begin{equation}
  \textstyle{\| K \|_1 = \sum_{i,j} |k_{i,j}|}
  \label{eq:l1_norm}
\end{equation}
A direct way to enforce sparsity in $K$ would be to add the $\ell_0$-norm constraint in (\ref{eq:SO_l0}) to problem (\ref{eq:optimization_problem}). However, while it is easier to assign $T$ for the $\ell_0$-norm constraint, it is difficult to choose it for the $\ell_1$ norm. Similarly, the weighted formulation (\ref{eq:SO_weighted}) introduces a hyperparameter $\lambda$ without physical interpretation and would require repeated, high‐dimensional cross‐validation with no convergence guarantee. Thus, we will use formulation (\ref{eq:SO_fc}).
\medskip\\
A sparse optimization problem can now be defined to sort variables according to their importance. This formulation is referred to as \textit{Structure optimization}.
\begin{subequations}
  \label{eq:opt_probl_structure}
  \begin{align}
    K = &\argmin_{K} \;\|\widetilde{K}\|_1 \label{eq:opt_probl_structure_objective}\\
    s.t.
    \quad &k_{i,j}\,\in\,[\,\underline{k}_{i,j},\,\overline{k}_{i,j}\,] \qquad
    \begin{matrix}
      \forall i = 1,\cdots,n_x\\
      \forall j = 1,\cdots,n_y
    \end{matrix}
    \label{eq:opt_probl_structure_var_limits}\\
    &\mathcal{G}(K) \leq 0
    \label{eq:opt_probl_structure_stability}\\
    &rms(\omega_z-\widehat{\omega}_z)\,<\,\text{UB}_{\omega_z} \label{eq:opt_probl_structure_ub_ome}\\
    &rms(v_x-\widehat{v}_x)\,\,<\,\text{UB}_{v_x} \label{eq:opt_probl_structure_ub_vx}
  \end{align}
\end{subequations}
In the objective function (\ref{eq:opt_probl_structure_objective}), $\widetilde{K}$ contains all the normalized parameters $\widetilde{k}_{i,j}$. Using the normalized parameters and not the original ones is crucial due to the varying orders of magnitude.\\
The normalization is computed as follows:
\begin{equation}
    \label{eq:parameter_normalization}
    \begin{split}
        \widetilde{k}_{i,j}\,=\,
        \begin{cases}
            \left\lvert\displaystyle{{{k}_{i,j}}\,/\,{\overline{k}_{i,j}}}\right\lvert,& \text{if } k_{i,j}\,>\,0\\
            \left\lvert\displaystyle{{k_{i,j}}\,/\,{{\underline{k}_{i,j}}}}\right\lvert,  & \text{otherwise}
        \end{cases}\\
    \forall\,i\,=\,1,\dots,n_x,\quad j\,=\,1,\dots,n_y\\
    \text{(}\,\text{by construction } \underline{k}_{i,j}\leq 0 \text{ and }\overline{k}_{i,j}\geq 0\,\text{)} \; .
    \end{split}
\end{equation}
Constraint (\ref{eq:opt_probl_structure_var_limits}) guarantees that the parameters are within the given optimization ranges and (\ref{eq:opt_probl_structure_stability}) represents the set of $K$ for which the filter is asymptotically stable. Constraints (\ref{eq:opt_probl_structure_ub_ome}) and (\ref{eq:opt_probl_structure_ub_vx}) guarantee a satisfactory performance on both lateral (\ref{eq:opt_probl_structure_ub_ome}) and longitudinal (\ref{eq:opt_probl_structure_ub_vx}) dynamics. Thus, their role is to mimic the 
terms in (\ref{eq:optimization_problem_objective}). The Upper Bounds on the rms errors $\text{UB}_{\omega_z}$ and $\text{UB}_{v_x}$ have physical meaning, hence they are much easier to select than the hyperparameter $\lambda$ in equation (\ref{eq:SO_weighted}). Unlike (\ref{eq:opt_probl_structure_var_limits}), these constraints are of the same nature as (\ref{eq:opt_probl_structure_stability}) and are unknown a priori.\\
In section \ref{sec:MBR}, the best optimizations had $rmse(v_x)\approx1.1 \; km/h$ and $rmse(\omega_z) \approx 0.6 \;deg/s$. By setting $\text{UB}(v_x)= 1.5\; km/h$ and $\text{UB}(\omega_z)= 1.5\;deg/s$ we can still have very good performance for $v_x$ and acceptable one for $\omega_z$.
\medskip\\
Although this reformulation introduces additional constraints while retaining the same number of decision variables, our goal is not the exact global optimum but rather a provably feasible solution that approaches optimality. By imposing sufficiently tight performance bounds (\ref{eq:opt_probl_structure_ub_vx}) and (\ref{eq:opt_probl_structure_ub_ome}), we ensure the quality of the solution. After solving (\ref{eq:opt_probl_structure})---even if suboptimally---the most sensitive parameters are identified, allowing a focused lower-dimensional optimization of the original problem (\ref{eq:optimization_problem}).
\\
The structure optimization casts a hard high-dimensional problem (\ref{eq:optimization_problem}) into a probably non-converging sparse optimization (\ref{eq:opt_probl_structure}), and is then followed by a reduced performance problem with a much greater probability of convergence to a better solution.

\subsection{Structure optimization procedure}
\label{subsec:structure_optimization_procedure}

\noindent
The result of the structure optimization (\ref{eq:opt_probl_structure}) are reported in Table \ref{table:struct_opt_result}.
\\
To prune the problem dimensionality, a threshold $\delta$ is introduced. All normalized $\tilde{k}_{i,j}$ below $\delta$ are removed. The remaining parameters are then optimized according to the original performance optimization problem (\ref{eq:optimization_problem}).
\\
In line with the three dimensionality‐reduction levels examined in Section \ref{sec:MBR}, we similarly adopt three pruning thresholds, $\delta$ = 0.05, 0.10 and 0.40. The resulting optimization classes will have 8, 6 and 4 parameters.

\begin{table}[]
  \centering
  \begin{minipage}{.45\columnwidth}
    \centering
    \begin{tabular}{| c | c | c |}
      \hline
      \textbf{Parameter} & $\boldsymbol{\widetilde{k}_{i,j}}$ & $\boldsymbol{k_{i,j}}$\T\B \\
      \hline \hline
      \textbf{$k_{\omega_{rr} \to \omega_{rr}}$} & 56.9\% & 0.85 \T\B \\
      \textbf{$k_{\omega_z \to \omega_{rr}}$} & 55.6\% & -0.83 \T\B \\
      \textbf{$k_{\omega_z \to \omega_z}$} & 42.2\% & 0.63 \T\B \\
      \textbf{$k_{\omega_{fl} \to \omega_{fl}}$} & 41.5\% & 0.62 \T\B \\
      \textbf{$k_{\omega_z \to v_x}$} & 35.2\% & 0.53 \T\B \\
      \textbf{$k_{\omega_{fl} \to v_x}$} & 26.2\% & 0.11 \T\B \\
      \hline
    \end{tabular}
  \end{minipage}
  \hspace{-3pt}
  \begin{minipage}{.45\columnwidth}
    \centering
    \begin{tabular}{| c | c | c |}
      \hline
      \textbf{Parameter} & $\boldsymbol{\widetilde{k}_{i,j}}$ & $\boldsymbol{k_{i,j}}$\T\B \\
      \hline \hline
      \textbf{$k_{\omega_{rr} \to \omega_{fl}}$} & 7.7\% & 0.058 \T\B \\
      \textbf{$k_{\omega_{fl} \to \omega_{rr}}$} & 5.5\% & -0.041 \T\B \\
      \textbf{$k_{\omega_{rr} \to \omega_z}$} & 3.8\% & 0.0023 \T\B \\
      \textbf{$k_{\omega_{fl} \to \omega_z}$} & 3.5\% & -0.0021\T\B \\
      \textbf{$k_{\omega_{rr} \to v_x}$} & 2.1\% & 0.0095 \T\B \\
      \textbf{$k_{\omega_z \to \omega_{fl}}$} & 2.0\% & 0.030 \T\B \\
      \hline
    \end{tabular}
  \end{minipage} 
  \\[5pt]
  \caption{Result of the structure optimization (\ref{eq:opt_probl_structure}).}
  \label{table:struct_opt_result}
  \vspace{-0.5cm}
\end{table}


\subsection{Statistical analysis}
\label{subsec:SDR_statistical_analysis}

\begin{figure}[]
  \centering
  \includegraphics[width=0.95\columnwidth]{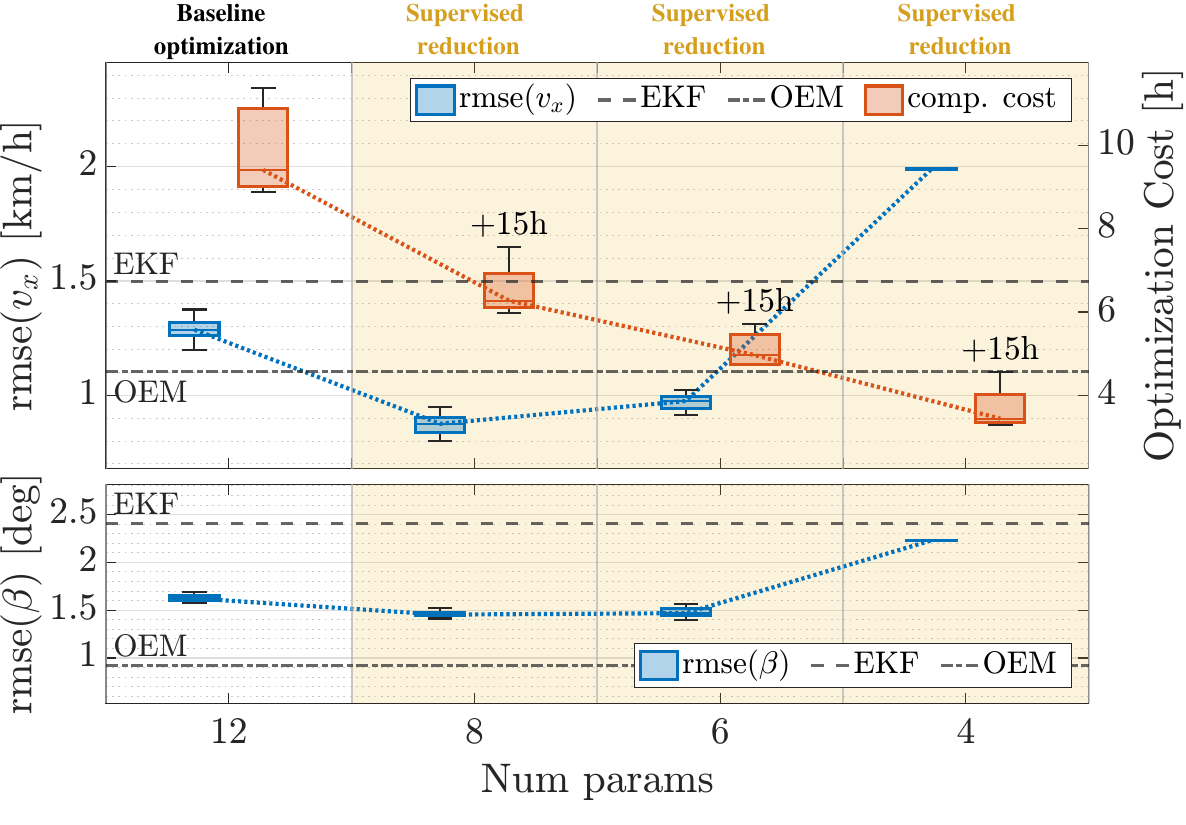}
  \vspace{-0.25cm}
  \caption{SDR: Validation A boxplot. Blue boxes report the distributions of rmse. Orange boxes represent the computational effort. Optimizations are sorted by descending dimensionality.}
  \label{fig:SDR_validation_boxplot}
\end{figure}

\noindent
Following the methodology of Section \ref{sec:MBR}, we performed 12 independent runs for each optimization class to quantify result variability. The optimal $K$ returned by each optimization was far from the variable range limits, with the only exception being the 4-dimensional case, where all three correlated parameters were saturated to their upper bound of 1.5.
\\
The boxplots containing the performance results for the validation lap (dataset A) are reported in Figure \ref{fig:SDR_validation_boxplot}. Optimizations are ordered from left to right in decreasing order of variables. We can draw similar conclusions to the MBR case (Figure \ref{fig:MBR_validation_boxplot}):
\begin{itemize}
  \item a clear parabolic trend in the median filter performance is visible (blue dotted line): the high-dimensional optimization is not converging to its global optimum, while the low-dimensional one  does not have all the degrees of freedom to converge to a well-performing solution;
  \item the variance in performance and computational times decrease with the number of parameters.
\end{itemize}
The results confirm that the dimensionality reduction is beneficial to improve the performance. The best solution (8-parameter one) is able to improve on average by 35\% with respect to the original optimization and outperforms by 21 \% the OEM filter. However, the performance improvement comes with an additional 15 hours of computational burden of the additional structure optimization, indicated with the ``+15h'' label. The structure optimization is more time‑consuming because it introduces the additional constraints (\ref{eq:opt_probl_structure_ub_vx}) and (\ref{eq:opt_probl_structure_ub_ome}), which must be estimated alongside the objective and incorporated into the acquisition function.\\
In addition, $\beta$ is slightly improved too, indicating that the correction is affecting the entire DT---even states that are not specifically included in the cost function.

\subsection{Additional validations}
\label{sec:MBR_extra_validations}

\begin{figure}[]
  \centering
  \hspace*{-0.08\columnwidth}
  \includegraphics[width=0.88\columnwidth]{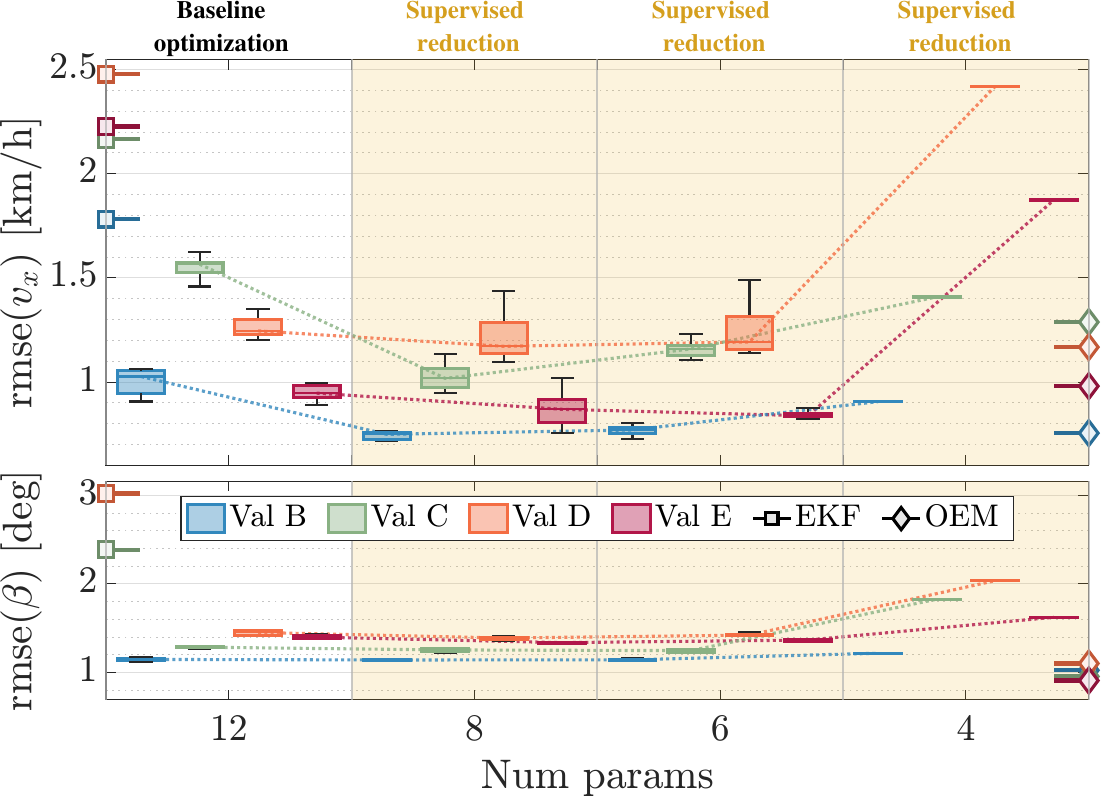}
  \vspace{-0.25cm}
  \caption{SDR: Additional validations boxplot for multiple datasets.}
  \label{fig:SDR_extraval_boxplot}
\end{figure}

\noindent
Figure \ref{fig:SDR_extraval_boxplot} reports the RMSE of $v_x$ and $\beta$ on datasets B–E for the SDR case. The parabolic shape of $v_x$ in Figure \ref{fig:SDR_validation_boxplot} for validation A reappears in the dotted curves of Figure \ref{fig:SDR_extraval_boxplot} for B-E. This effect is most pronounced in $v_x$ and less visible in $\beta$, due to axis scaling. It shows that the 6‑ and 8‑dimensional filters outperform the 12‑d baseline, whereas the 4‑d model remains too sparse to correct the DT effectively. Notably, the 6- and 8‑parameter SDR observers now outperform the OEM $v_x$ filters by 4\%, 21\%, and 15\% on B, C, and E, respectively, and matching the OEM on D. Moreover, SDR filters consistently outperform the MBR ones in every validation dataset.

\section{Unsupervised dimensionality reduction}
\label{sec:UDR}

\begin{figure}[]
  \centering
  \includegraphics[width=0.95\columnwidth]{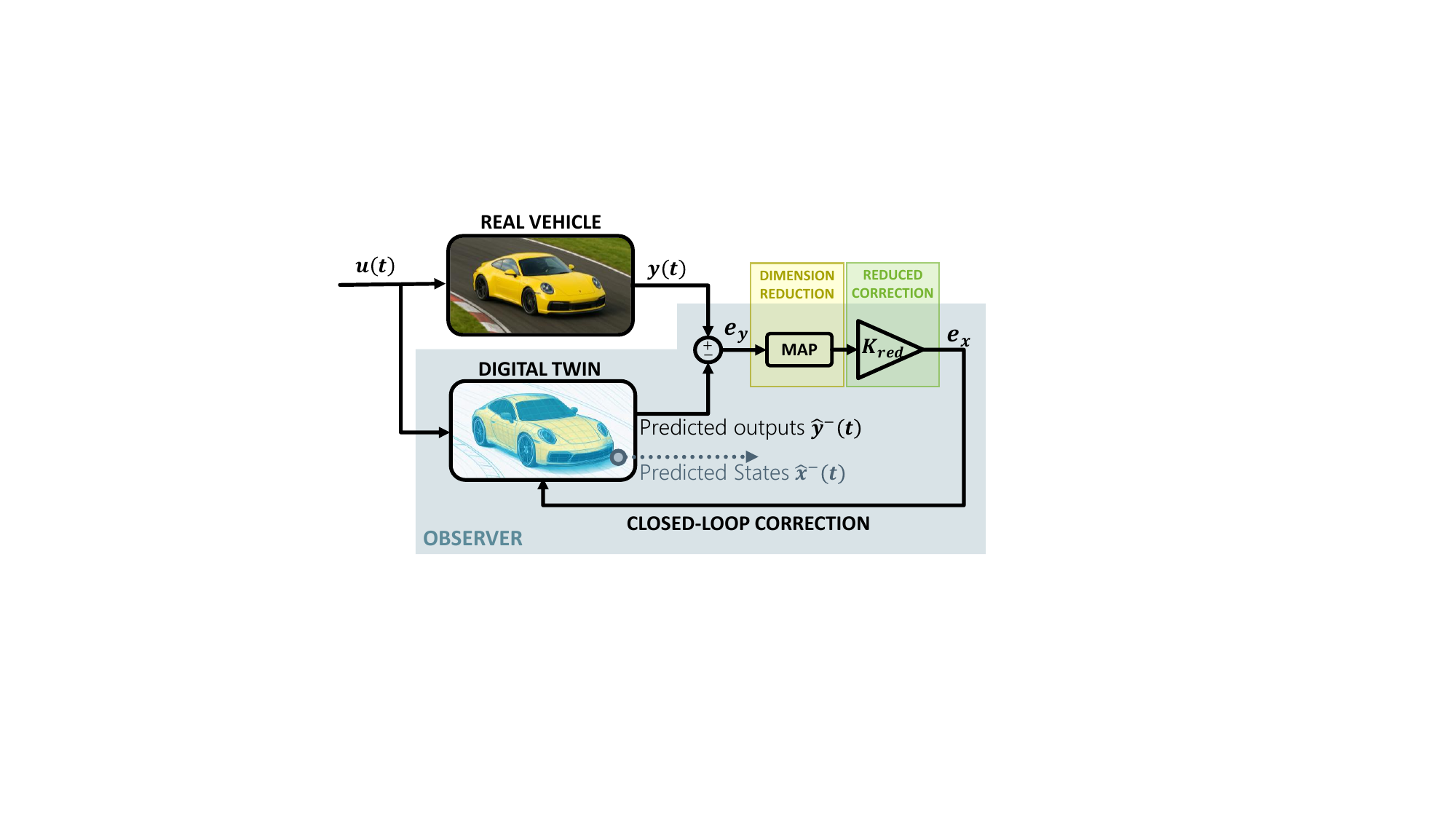}
  \caption{Twin-in-the-Loop Observer architecture with input reduction.}
  \label{fig:TiL_scheme_PCA}
\end{figure}

\noindent
The complexity of tuning a full‑scale TiL correction matrix grows combinatorially with each added state or output, rendering impractical to use BO while when the number of corrected states or employed measurements is higher.\\
Our earlier work \cite{TiL_filtering_PCA} addressed precisely this challenge by reducing a 16‑variable correction matrix---arising from four states and four outputs---via Principal Component Analysis (PCA). This Unsupervised DR approach inserts a linear map (Figure \ref{fig:TiL_scheme_PCA}) before the correction stage to project the output error $e_y$ into a lower dimensional $\tilde{e}_y$ subspace. The dimensionality reduction matrix is not learned on the measured $e_y$, which would depend on the correction matrix $K$ and would lead to unreliable results \cite{TiL_filtering_PCA}, but on the measured outputs $y$, and can be directly applied to $e_y$ because of its linearity.
The original gain $K$ can thus be replaced with a reduced‑input gain $K_{red}$, whose input dimension $\tilde{n}_y$ equals the number of retained principal components. Since this reduction is agnostic to the target loss, it will be referred to as ``Unsupervised DR''.\\
In \cite{TiL_filtering_PCA}, the authors show that PCA retains 99,98\% of the cumulated power of the singular values while projecting the original 4 measurements into a 3 dimensional subspace, casting the 16-dimensional $K$ matrix into a 12-dimensional $K_{red}$ matrix.\\
One of the key difficulty in UDR is that the optimization ranges in (\ref{eq:optimization_problem}) refer to $K$, yet after feature extraction they must be stated for $K_{red}$. Under a linear extraction map, Theorem I in \cite{TiL_filtering_PCA} provides a sufficient (but not necessary) condition to convert $K$ bounds into $K_{red}$ bounds.

\subsection{Statistical analysis}
\label{subsec:UDR_statistical_analysis}

\noindent
Figure \ref{fig:UDR_validation_boxplot} reports the boxplots that compare the 16-dimensional optimization reduced to 12 using PCA and the native 12-dimensional benchmark.
\\
Note that PCA is able to get better results on average for $v_x$ (blue line), but at the cost of an increased variance and doubled computational effort. These drawbacks are not surprising: Theorem 1 \cite{TiL_filtering_PCA} is only sufficient, hence the limits in $K_{red}$ are intrinsically larger than their $K$ counterpart. This introduces more complex unstable regions, which increase the number of unfeasible BO iterations, thus, the computational effort. These results highlight even further the importance of the choice of the optimization variable ranges.

\subsection{Combined SDR and UDR approach}
\label{subsec:SUDR_combined_approach}

\begin{figure}[]
  \centering
  \includegraphics[width=0.95\columnwidth]{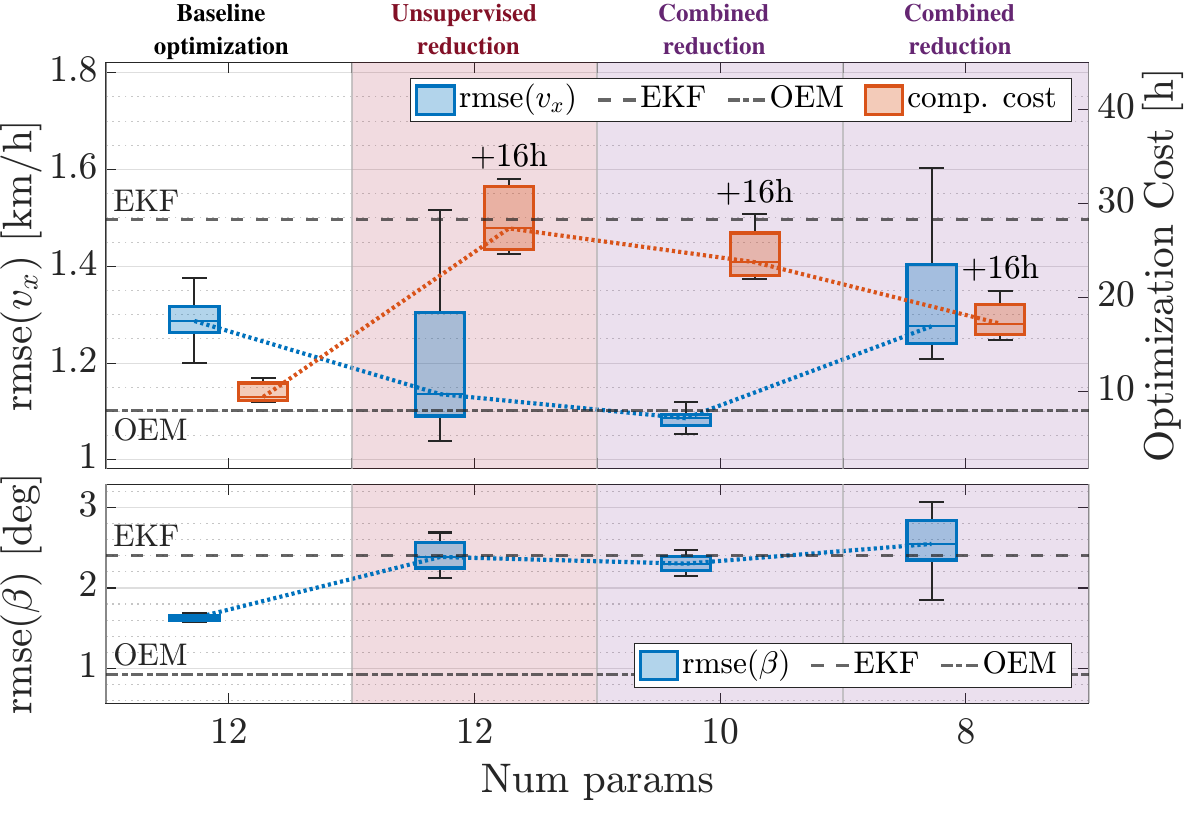}
  \vspace{-0.25cm}
  \caption{UDR and combined SDR + UDR approach: Validation A boxplot. Blue boxes report the distributions of rmse. Orange boxes represent the computational effort. Optimizations are sorted by descending dimensionality.}
  \label{fig:UDR_validation_boxplot}
\end{figure}

\noindent
We now investigate a hybrid dimensionality‐reduction strategy that combines the supervised and unsupervised methods. Specifically, we apply the very same structure‑optimization routine, previously defined for $K$, directly to the PCA‑reduced gain matrix $K_{red}$. Beginning with a 12‑dimensional problem (down from 16), we impose sparsity thresholds $\delta$: 0.07 and 0.1, which respectively correspond to removing 2 and 4 variables. More severe reductions (i.e. for $\delta > 0.1$) proved to be destabilize the filter. All UDR configurations show reduced $\beta$ estimation accuracy, likely due to degraded lateral‑tracking performance in the reduced model.
\\
The statistical results obtained by repeating each optimization type 12 times are reported in Figure \ref{fig:UDR_validation_boxplot} (purple areas). The parabolic trend present for $v_x$ in \ref{fig:SDR_validation_boxplot} is also evident here. Due to the complexity of the feature extraction layer, the variance of the lower-dimensional case is not the smallest. This is even more evident if we remove more parameters, since the observer becomes unstable.\\
The computational times of all the UDR cases increase considerably with respect to both the baseline and the MBR and SDR cases. This is due to the fact that the larger optimization ranges lead to a much higher number of unfeasible iterations---about as many as the feasible ones---hence doubling the number of overall sampled points.

\begin{figure*}[]
  \centering
  \includegraphics[width=0.9\textwidth]{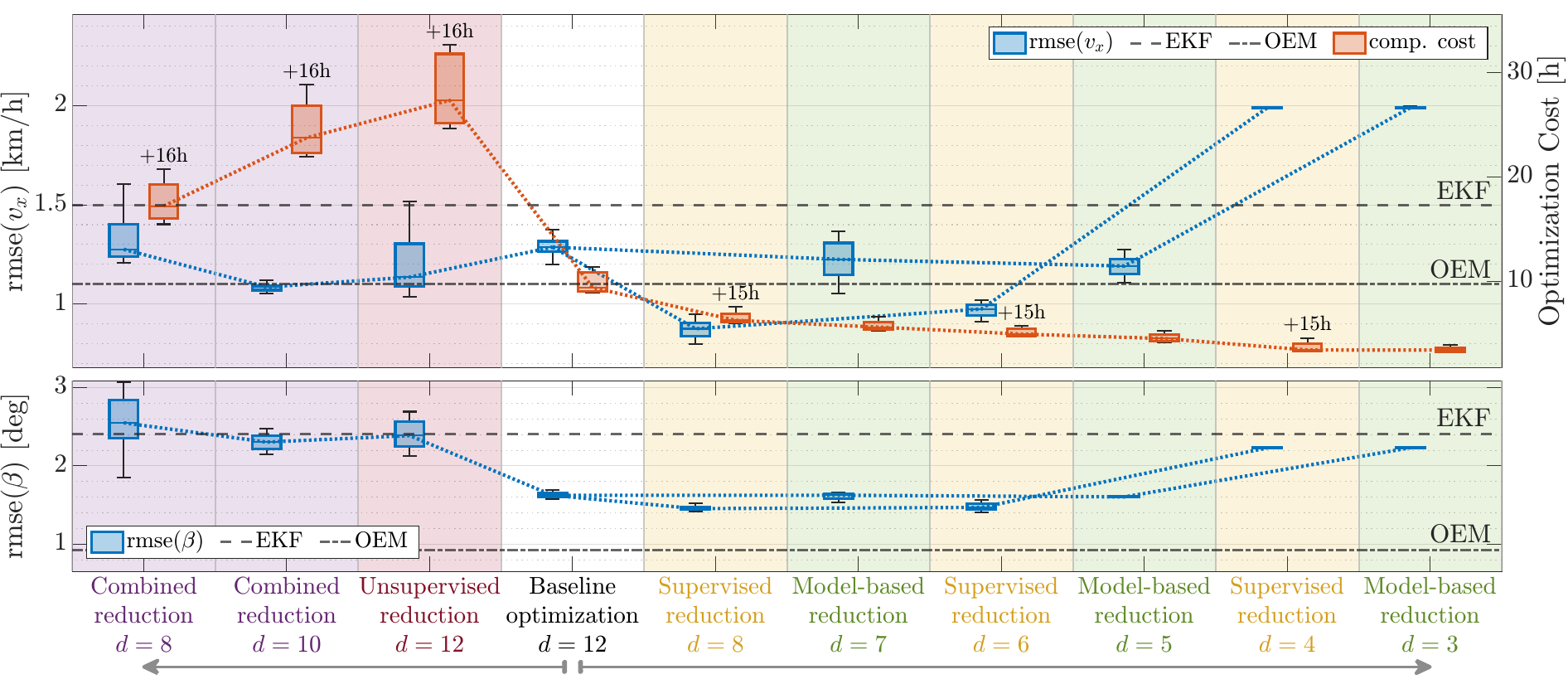}
  \caption{Result comparison boxplot. This figure summarizes Figures \ref{fig:MBR_validation_boxplot}, \ref{fig:SDR_validation_boxplot} and \ref{fig:UDR_validation_boxplot}.}
  \label{fig:final_boxplot}
\end{figure*}

\subsection{Overall comparison}

\noindent
Let us now compare the three techniques (MBR, SDR, and UDR). Figure \ref{fig:final_boxplot} reports all the statistical analyses seen in Figures \ref{fig:MBR_validation_boxplot}, \ref{fig:SDR_validation_boxplot} and \ref{fig:UDR_validation_boxplot}. The original 12‑parameter solution (white background) is compared with MBR (green) and SDR (yellow) on the right, and with UDR (red) and combined UDR+SDR (purple) on the left. All methods exhibit a parabolic median trend: reducing dimensionality improves convergence up to a point, after which performance deteriorates. SDR achieves the lowest errors, outperforming both the EKF and the OEM filter, confirming that data‐driven reduction (SDR) surpasses prior‐knowledge pruning (MBR), which was adopted in \cite{TiL_Riva}. The most aggressive reductions (3–4 parameters) produce underperforming observers.\\
Although UDR variants may occasionally surpass the full‐dimensional baseline, they exhibit substantially greater variability and a degraded accuracy in lateral dynamics ($\beta$).\\
Computational cost decreases as dimensionality is reduced with SDR and MBR (right), but approximately doubles for PCA‑based reductions (left), due to the additional complexity of the PCA layer and the larger optimization ranges.

\medskip
\noindent
We can summarize the results as follows: if the optimizer cannot converge within a reasonable time, unsupervised dimensionality reduction (UDR) is the only viable option; once the problem becomes tractable but still involves a high-dimensional search space, applying dimensionality reduction---ideally supervised (SDR)---yields better convergence and accuracy. In the latter case, the choice of the SDR sparsity‑imposing threshold $\delta$ depends on multiple factors, such as the presence of a clear knee in the normalized variables importance profile (Table \ref{table:struct_opt_result}), the optimizer’s capability to handle the resulting problem, and the computational time available. The ultimate choice between UDR, SDR or the full-dimensional problem is inherently dependent on both problem complexity and the solver capabilities.


\section{Conclusions}
\label{sec:conclusions}

\noindent
In this study, we explored two dimensionality reduction techniques aimed at addressing non-converging large-scale black-box optimization problems. In Section \ref{sec:MBR}, we introduced a model-based benchmark (MBR), while Sections \ref{sec:SDR} and \ref{sec:UDR} presented two distinct reduction approaches: SDR and UDR, compared in Section \ref{subsec:SUDR_combined_approach}. All reduction methods were evaluated in the context of performance optimization for a high-performing vehicle on a race track.
The heuristic proposed in Section \ref{subsec:optimization_ranges_heuristic} facilitated the identification of parameters required for optimization (\ref{eq:optimization_problem}) with minimal prior system knowledge.

\noindent
Future endeavors will focus extensive testing of estimator performance, detailed analysis of their stability, and exploration of different optimization methodologies.


\section*{Acknowledgments}
This paper is supported by FAIR (Future Artificial Intelligence Research) project, funded by the NextGenerationEU program within the PNRR-PE-AI scheme (M4C2, Investment 1.3, Line on Artificial Intelligence), and by the Italian Ministry of Enterprises and Made in Italy in the framework of the project 4DDS (4D Drone Swarms) under grant no. F/310097/01-04/X56.

\bibliography{IEEEabrv, paper_bib.bib}

\begin{thebibliography}{10}
\providecommand{\url}[1]{#1}
\csname url@samestyle\endcsname
\providecommand{\newblock}{\relax}
\providecommand{\bibinfo}[2]{#2}
\providecommand{\BIBentrySTDinterwordspacing}{\spaceskip=0pt\relax}
\providecommand{\BIBentryALTinterwordstretchfactor}{4}
\providecommand{\BIBentryALTinterwordspacing}{\spaceskip=\fontdimen2\font plus
\BIBentryALTinterwordstretchfactor\fontdimen3\font minus \fontdimen4\font\relax}
\providecommand{\BIBforeignlanguage}[2]{{%
\expandafter\ifx\csname l@#1\endcsname\relax
\typeout{** WARNING: IEEEtran.bst: No hyphenation pattern has been}%
\typeout{** loaded for the language `#1'. Using the pattern for}%
\typeout{** the default language instead.}%
\else
\language=\csname l@#1\endcsname
\fi
#2}}
\providecommand{\BIBdecl}{\relax}
\BIBdecl

\bibitem{vehicle_filters}
K.~B. Singh, M.~A. Arat, and S.~Taheri, ``Literature review and fundamental approaches for vehicle and tire state estimation,'' \emph{Vehicle System Dynamics}, vol.~57, no.~11, pp. 1643--1665, 2019.

\bibitem{TiL_Riva}
G.~Riva, S.~Formentin, M.~Corno, and S.~M. Savaresi, ``Twin-in-the-loop state estimation for vehicle dynamics control: Theory and experiments,'' \emph{IFAC Journal of Systems and Control}, vol.~29, p. 100274, 2024.

\bibitem{TiL_control_long}
F.~Dettù, S.~Formentin, and S.~M. Savaresi, ``The twin-in-the-loop approach for vehicle dynamics control,'' \emph{IEEE/ASME Transactions on Mechatronics}, pp. 1--12, 2023.

\bibitem{TiL_control_lat}
F.~Dettù, G.~Delcaro, S.~Formentin, S.~Varisco, and S.~M. Savaresi, ``Optimization tools for twin-in-the-loop vehicle control design: analysis and yaw-rate tracking case study,'' \emph{European Journal of Control}, vol.~77, p. 100998, 2024.

\bibitem{TiL_filtering_params}
{Dettù, Federico and Formentin, Simone and Savaresi, Sergio Matteo}, ``Joint vehicle state and parameters estimation via twin-in-the-loop observers,'' \emph{Vehicle System Dynamics}, vol.~0, no.~0, pp. 1--27, 2023.

\bibitem{TiL_filtering_robustness}
F.~Dettù, S.~Formentin, and S.~M. Savaresi, ``Robust tuning of twin-in-the-loop vehicle dynamics controls via randomized optimization,'' in \emph{Proc. 22nd IFAC World Congress}, 2023.

\bibitem{TiL_filtering_PCA}
G.~Delcaro, F.~Dettù, S.~Formentin, and S.~M. Savaresi, ``Dealing with the curse of dimensionality in twin-in-the-loop observer design,'' in \emph{Proc. 22nd IFAC World Congress}, 2023.

\bibitem{whitebox_unified_observer}
A.~J. Rodríguez, E.~Sanjurjo, R.~Pastorino, and M.~Ángel Naya, ``State, parameter and input observers based on multibody models and kalman filters for vehicle dynamics,'' \emph{Mechanical Systems and Signal Processing}, vol. 155, p. 107544, 2021.

\bibitem{whitebox_switching_observer}
L.-Y. Hsu and T.-L. Chen, ``Vehicle full-state estimation and prediction system using state observers,'' \emph{IEEE Transactions on Vehicular Technology}, vol.~58, no.~6, pp. 2651--2662, 2009.

\bibitem{BO}
B.~Shahriari, K.~Swersky, Z.~Wang, R.~P. Adams, and N.~de~Freitas, ``Taking the human out of the loop: A review of bayesian optimization,'' \emph{Proceedings of the IEEE}, vol. 104, no.~1, pp. 148--175, 2016.

\bibitem{scenario_approach}
M.~C. Campi, S.~Garatti, and M.~Prandini, ``The scenario approach for systems and control design,'' \emph{Annual Reviews in Control}, vol.~33, no.~2, pp. 149--157, 2009.

\bibitem{BO_frazier}
\BIBentryALTinterwordspacing
P.~I. Frazier, ``A tutorial on bayesian optimization,'' 2018. [Online]. Available: \url{https://arxiv.org/abs/1807.02811}
\BIBentrySTDinterwordspacing

\bibitem{BO_cod1}
D.~Eriksson and M.~Jankowiak, ``High-dimensional {Bayesian} optimization with sparse axis-aligned subspaces,'' in \emph{Proceedings of the Thirty-Seventh Conference on Uncertainty in Artificial Intelligence}, vol. 161.\hskip 1em plus 0.5em minus 0.4em\relax PMLR, 2021, pp. 493--503.

\bibitem{BO_cod2}
R.~Moriconi, M.~P. Deisenroth, and K.~Sesh~Kumar, ``High-dimensional bayesian optimization using low-dimensional feature spaces,'' \emph{Machine Learning}, vol. 109, pp. 1925--1943, 2020.

\bibitem{GP_and_BO}
M.~Binois and N.~Wycoff, ``A survey on high-dimensional gaussian process modeling with application to bayesian optimization,'' \emph{ACM Transactions on Evolutionary Learning and Optimization}, vol.~2, no.~2, pp. 1--26, 2022.

\bibitem{SMGO}
L.~Sabug, F.~Ruiz, and L.~Fagiano, ``{SMGO-$\Delta$}: Balancing caution and reward in global optimization with black-box constraints,'' \emph{Information Sciences}, vol. 605, pp. 15--42, 2022.

\bibitem{DR_1}
W.~Jia, M.~Sun, J.~Lian, and S.~Hou, ``Feature dimensionality reduction: a review,'' \emph{Complex \& Intelligent Systems}, vol.~8, no.~3, pp. 2663--2693, 2022.

\bibitem{LDR}
J.~P. Cunningham and Z.~Ghahramani, ``Linear dimensionality reduction: Survey, insights, and generalizations,'' \emph{The Journal of Machine Learning Research}, vol.~16, no.~1, pp. 2859--2900, 2015.

\bibitem{highdim_BO_1}
Z.~Wang, F.~Hutter, M.~Zoghi, D.~Matheson, and N.~De~Feitas, ``Bayesian optimization in a billion dimensions via random embeddings,'' \emph{Journal of Artificial Intelligence Research}, vol.~55, pp. 361--387, 2016.

\bibitem{highdim_BO_2}
\BIBentryALTinterwordspacing
B.~Chen, R.~Castro, and A.~Krause, ``Joint optimization and variable selection of high-dimensional gaussian processes,'' 2012. [Online]. Available: \url{https://arxiv.org/abs/1206.6396}
\BIBentrySTDinterwordspacing

\bibitem{highdim_BO_3}
A.~Nayebi, A.~Munteanu, and M.~Poloczek, ``A framework for {B}ayesian optimization in embedded subspaces,'' in \emph{Proceedings of the 36th International Conference on Machine Learning}, ser. Proceedings of Machine Learning Research, vol.~97.\hskip 1em plus 0.5em minus 0.4em\relax PMLR, 2019, pp. 4752--4761.

\bibitem{SO_wright}
S.~Wright, ``Sparse optimization methods,'' in \emph{Conference on Advanced Methods and Perspectives in Nonlinear Optimization and Control}, 2010.

\bibitem{candes06}
E.~J. Candès, J.~K. Romberg, and T.~Tao, ``Stable signal recovery from incomplete and inaccurate measurements,'' \emph{Communications on Pure and Applied Mathematics}, vol.~59, no.~8, pp. 1207--1223, 2006.

\bibitem{matrix_completion}
\BIBentryALTinterwordspacing
E.~Cand\`{e}s and B.~Recht, ``Exact matrix completion via convex optimization,'' \emph{Commun. ACM}, vol.~55, no.~6, p. 111–119, jun 2012. [Online]. Available: \url{https://doi.org/10.1145/2184319.2184343}
\BIBentrySTDinterwordspacing

\bibitem{matrix_completion_with_noise}
E.~J. Candes and Y.~Plan, ``Matrix completion with noise,'' \emph{Proceedings of the IEEE}, vol.~98, no.~6, pp. 925--936, 2010.

\bibitem{zerothorderopt}
S.~Liu, P.-Y. Chen, B.~Kailkhura, G.~Zhang, A.~O. Hero~III, and P.~K. Varshney, ``A primer on zeroth-order optimization in signal processing and machine learning: Principals, recent advances, and applications,'' \emph{IEEE Signal Processing Magazine}, vol.~37, no.~5, pp. 43--54, 2020.

\bibitem{surrogate_functions_optimization}
N.~V. Queipo, R.~T. Haftka, W.~Shyy, T.~Goel, R.~Vaidyanathan, and P.~{Kevin Tucker}, ``Surrogate-based analysis and optimization,'' \emph{Progress in Aerospace Sciences}, vol.~41, no.~1, pp. 1--28, 2005.

\bibitem{ego}
D.~R. Jones, M.~Schonlau, and W.~J. Welch, ``Efficient global optimization of expensive black-box functions,'' \emph{Journal of Global optimization}, vol.~13, pp. 455--492, 1998.

\bibitem{GP_book}
C.~K. Williams and C.~E. Rasmussen, \emph{Gaussian processes for machine learning}.\hskip 1em plus 0.5em minus 0.4em\relax MIT press Cambridge, MA, 2006, vol.~2, no.~3.

\bibitem{ARD}
D.~J. MacKay and R.~M. Neal, ``Automatic relevance determination for neural networks,'' in \emph{Technical Report in preparation.}\hskip 1em plus 0.5em minus 0.4em\relax Cambridge University, 1994.

\bibitem{coupled_constraint}
M.~Khosravi, C.~König, M.~Maier, R.~S. Smith, J.~Lygeros, and A.~Rupenyan, ``Safety-aware cascade controller tuning using constrained bayesian optimization,'' \emph{IEEE Transactions on Industrial Electronics}, vol.~70, no.~2, pp. 2128--2138, 2023.

\bibitem{BO_with_constraints}
\BIBentryALTinterwordspacing
M.~A. Gelbart, J.~Snoek, and R.~P. Adams, ``Bayesian optimization with unknown constraints,'' 2014. [Online]. Available: \url{https://arxiv.org/abs/1403.5607}
\BIBentrySTDinterwordspacing

\bibitem{parallelBO}
\BIBentryALTinterwordspacing
D.~Ginsbourger, R.~Le~Riche, and L.~Carraro, ``{A Multi-points Criterion for Deterministic Parallel Global Optimization based on Gaussian Processes},'' Tech. Rep., Mar. 2008. [Online]. Available: \url{https://hal.science/hal-00260579}
\BIBentrySTDinterwordspacing

\bibitem{survey_Puscul}
D.~Puščul, C.~Lex, M.~Vignati, and L.~Shao, ``A literature survey on sideslip angle estimation using vehicle dynamics based methods,'' \emph{IEEE Access}, vol.~PP, pp. 1--1, 01 2024.

\bibitem{survey_Chindamo}
D.~Chindamo, B.~Lenzo, and M.~Gadola, ``On the vehicle sideslip angle estimation: a literature review of methods, models, and innovations,'' \emph{applied sciences}, vol.~8, no.~3, p. 355, 2018.

\bibitem{ekf_reina}
A.~Leanza, G.~Reina, and G.~Mantriota, ``Model-based observers for vehicle dynamics and tyre force prediction,'' \emph{Vehicle System Dynamics}, vol.~60, 05 2021.

\bibitem{SO}
Z.~Zhang, Y.~Xu, J.~Yang, X.~Li, and D.~Zhang, ``A survey of sparse representation: Algorithms and applications,'' \emph{IEEE Access}, vol.~3, pp. 490--530, 2015.

\end{thebibliography}
\bibliographystyle{IEEEtran}


\newlength{\biovertnegspace}      
\setlength{\biovertnegspace}{-0.75cm}

\newpage
\begin{IEEEbiography}[{\includegraphics[width=1in,height=1.25in,clip,keepaspectratio]{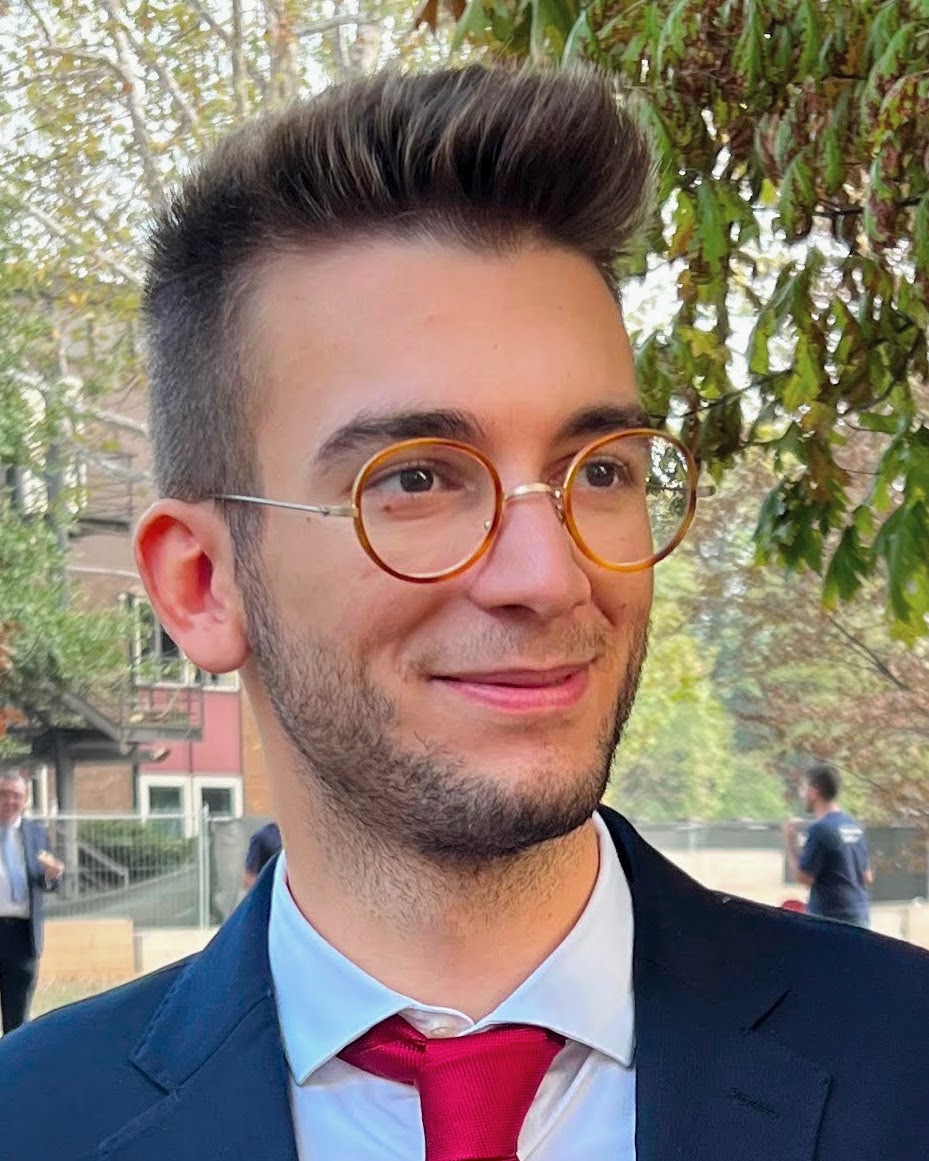}}]{Giacomo Delcaro}
received the B.Sc. and M.Sc. degrees (cum laude) in Automation and Control Engineering at Politecnico di Milano in 2020 and 2022. In 2020 he joined the mOve Research Group and spent four months in Indianapolis (USA) for the IndyAutonomousChallenge competition. In November 2022 he began his Ph.D. in Systems and Control, at the Department of Electronics, Computer Science and Bioengineering (DEIB) of Politecnico di Milano. His research includes innovative data-driven estimation and control methods for TiL systems.
\end{IEEEbiography}

\vspace{\biovertnegspace}

\begin{IEEEbiography}[{\includegraphics[width=1in,height=1.25in,clip,keepaspectratio]{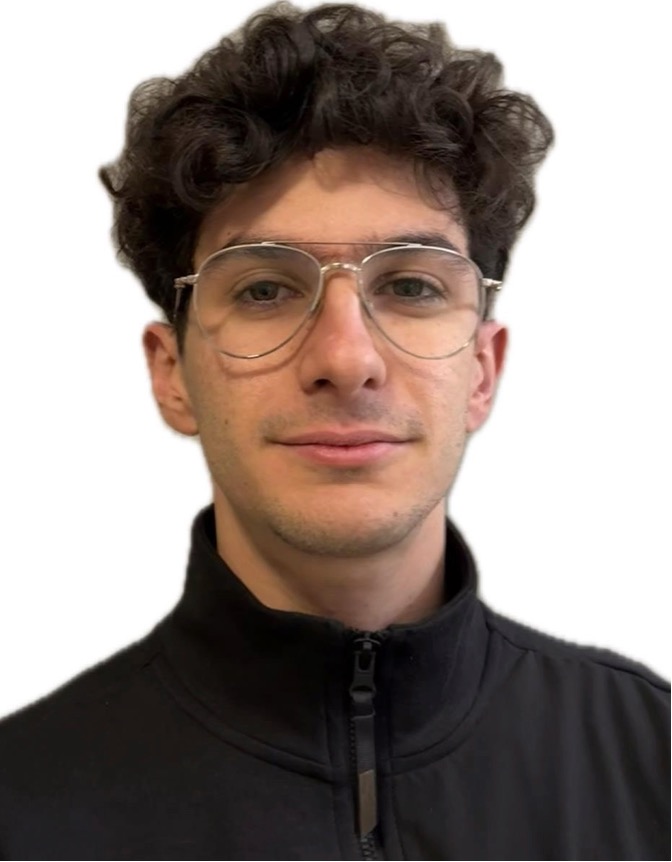}}]{Riccardo Poli}
received the B.Sc. degree and the M.Sc. degrees in automation and control engineering at Politecnico di Milano in 2021 and 2023. In March 2023, he joined the mOve Research Group as a M.Sc. student and became part of the PoliMOVE Autonomous Racing Team. In December 2023 he began his Ph.D. in Systems and Control, at the Department of Electronics, Computer Science and Bioengineering (DEIB) of Politecnico di Milano. His research interests include autonomous racecar localization and state estimation.
\end{IEEEbiography}

\vspace{\biovertnegspace}

\begin{IEEEbiography}[{\includegraphics[width=1in,height=1.25in,clip,keepaspectratio]{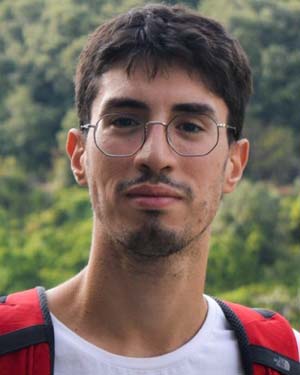}}]{Federico Dettù}
received the B.Sc. degree and the M.Sc. degree (cum laude) in automation and control engineering from Politecnico di Milano in 2017 and 2020. In 2020, he was a Visiting Researcher with Stanford University, Stanford, USA. In November 2020, he joined the mOve Research Group as a Ph.D. Student in Systems and Control, at the Department of Electronics, Computer Sciences and Bioengineering (DEIB) of Politecnico di Milano. His research interests include data-based estimation and control approaches for automotive systems.
\end{IEEEbiography}

\vspace{\biovertnegspace}

\begin{IEEEbiography}[{\includegraphics[width=1in,height=1.25in,clip,keepaspectratio]{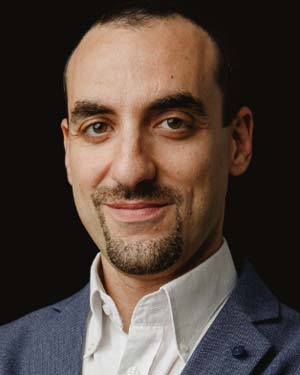}}]{Simone Formentin}
was born in Legnano, Italy, in 1984. He received the B.Sc. and M.Sc. degrees (cum laude) in automation and control engineering from Politecnico di Milano in 2006 and 2008. He received the Ph.D. degree (cum laude) in information technology within a joint program between Politecnico di Milano and Johannes Kepler University of Linz in 2012. After that, he held two postdoctoral appointments with the Swiss Federal Institute of Technology of Lausanne (EPFL) and the University of Bergamo. Since 2014, he has been with Politecnico di Milano, first as an Assistant Professor, then as an Associate Professor. His research interests include system identification and data-driven control with a focus on automotive and financial applications. Dr. Formentin is the Chair of the IEEE Technical Committee on System Identification and Adaptive Control, the social media representative of the IFAC Technical Committee on Robust Control, and a Member of the IFAC Technical Committee on Modelling, Identification and Signal Processing. He is an Associate Editor for Automatica and the European Journal of Control.
\end{IEEEbiography}

\vspace{\biovertnegspace}

\begin{IEEEbiography}[{\includegraphics[width=1in,height=1.25in,clip,keepaspectratio]{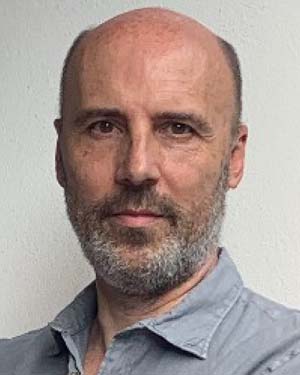}}]{Sergio Matteo Savaresi}
(Senior Member, IEEE) received the M.Sc. degree in electrical engineering from Politecnico di Milano in 1992, and the M.Sc. degree in applied mathematics from Catholic University, Brescia, Italy, in 2000, and the Ph.D. degree in systems and control engineering from Politecnico di Milano, in 1996. After his Ph.D., he was a Management Consultant with McKinsey \& Company, Milan, Italy. Since 2006, he has been a Full Professor in automatic control at Politecnico di Milano, where he is Deputy Director and Chair of the Systems and Control Section of the Department of Electronics, Computer Sciences and Bioengineering (DEIB). He has been a Manager and Technical Leader of more than 400 research projects in cooperation with private companies. He is the Co-Founder of nine high-tech startup companies. He is the author of more than 500 scientific publications. His main research interests include vehicle control, automotive systems, data analysis and system identification, with a special focus on smart mobility.
\end{IEEEbiography}

\vfill

\end{document}